\documentclass[twocol]{ametsocV5_arxiv}

\usepackage{amsmath,amsfonts,amssymb,bm}
\usepackage{mathptmx}
\usepackage{newtxtext}
\usepackage{newtxmath}
\usepackage{bbm}
\usepackage{graphicx}
\usepackage{cjhebrew}
\usepackage{xr}
\makeatletter
\patchcmd{\Ginclude@eps}{"#1"}{#1}{}{}
\makeatother

\makeatletter
\newcommand*{\addFileDependency}[1]{
  \typeout{(#1)}
  \@addtofilelist{#1}
  \IfFileExists{#1}{}{\typeout{No file #1.}}
}
\makeatother

\newcommand{\pr}{\mathbb{P}}
\newcommand{\ex}{\mathbb{E}}
\newcommand{\gcal}{\mathcal{G}}

\newcommand{\lcal}{\mathcal{L}}

\newcommand{\acal}{\mathcal{A}}
\newcommand{\ecal}{\mathcal{E}}

\newcommand{\tr}{^\top}

\newcommand{\inv}{^{-1}}

\newcommand{\vx}{\mathbf{x}}
\newcommand{\bx}{\mathbf{X}}
\newcommand{\bj}{\mathbf{J}}

\newcommand{\vn}{\mathbf{n}}

\newcommand{\vb}{\mathbf{b}}
\newcommand{\va}{\mathbf{a}}

\newcommand{\ov}{\overline}
\newcommand{\re}{\text{Re}}
\newcommand{\im}{\text{Im}}
\newcommand{\fn}{\frac{f_0^2}{N^2}}

\newcommand{\pder}[2]{\frac{\partial{#1}}{\partial{#2}}}

\newcommand{\eps}{\varepsilon}
\newcommand{\bmt}{\begin{bmatrix}}
\newcommand{\emt}{\end{bmatrix}}

\newcommand{\R}{\mathbb{R}}

\newcommand{\qp}{q^+}
\newcommand{\qm}{q^-}
\newcommand{\tbd}{\eta^+}

\newcommand{\gramps}{\Gamma}

\title{Data-driven transition path analysis yields a statistical understanding of sudden stratospheric warming events in an idealized model}

\authors{Justin Finkel\correspondingauthor{Justin Finkel, ju26596@mit.edu}}
\affiliation{Department of Earth, Atmospheric and Planetary Sciences, Massachusetts Institute of Technology}

\extraauthor{Robert J. Webber}
\extraaffil{Department of Computing and Mathematical Sciences, California Institute of Technology}

\extraauthor{Edwin P. Gerber}
\extraaffil{Courant Institute of Mathematical Sciences, New York University}

\extraauthor{Dorian S. Abbot}
\extraaffil{Department of the Geophysical Sciences, University of Chicago}

\extraauthor{Jonathan Weare}
\extraaffil{Courant Institute of Mathematical Sciences, New York University}

\abstract{
	Atmospheric regime transitions are highly impactful as drivers of extreme weather events, but pose two formidable modeling challenges: predicting the next event (weather forecasting), and characterizing the statistics of events of a given severity (the risk climatology). Each event has a different duration and spatial structure, making it hard to define an objective ``average event.'' We argue here that transition path theory (TPT), a stochastic process framework, is an appropriate tool for the task. We demonstrate TPT's capacities on a wave-mean flow model of sudden stratospheric warmings (SSWs) developed by \cite{holton_mass}, which is idealized enough for transparent TPT analysis but complex enough to demonstrate computational scalability. Whereas a recent article \citep{Finkel2021learning} studied near-term SSW predictability, the present article uses TPT to link predictability to long-term SSW frequency. This requires not only forecasting forward in time from an initial condition, but also \emph{backward in time} to assess the probability of the initial conditions themselves. TPT enables one to condition the dynamics on the regime transition occurring, and thus visualize its physical drivers with a vector field called the \emph{reactive current}. The reactive current shows that before an SSW, dissipation and stochastic forcing drive a slow decay of vortex strength at lower altitudes. The response of upper-level winds is late and sudden, occurring only after the transition is almost complete from a probabilistic point of view. This case study demonstrates that TPT quantities, visualized in a space of physically meaningful variables, can help one understand the dynamics of regime transitions. 
}

\begin{document}

\maketitle

%
%
%

%

\section{Introduction}

Many features of the atmosphere-ocean system's large-scale variability can be viewed as transitions between qualitatively different regimes. Examples include blocking, monsoons, El Ni{\~n}o, and Sudden Stratospheric Warming events (SSWs, the subject of this paper), all of which are associated with extreme weather. From a scientific perspective, regime transitions are handles by which to probe the climate's nonlinear, non-equilibrium dynamics. They expose novel physics and push us to qualitatively expand our physical understanding. From a human perspective, these relatively rare anomalies pose major societal challenges \citep{Lesk2016influence,Kron2019changes}, especially with a changing climate and increasing reliance on weather-susceptible infrastructure \citep[e.g.,][]{Mann2017influence,Frame2020attribution}. 

Regime transitions are used as benchmarks for model development across a hierarchy, from state-of-the-art Earth system models with billions of variables \citep[e.g.,][]{Stephenson2008extreme,Lengaigne2010contrasting,Vitart2018s2s} to conceptual low-order models with fewer than ten variables \citep[e.g.,][]{Charney1979,Timmermann2003nonlinear,ruz,Crommelin2004,Thual2016simple}.
In~\cite{Finkel2021learning}, we addressed near term forecasting of regime transitions in the context of an idealized sudden stratospheric warming (SSW) model constructed by \cite{holton_mass}, which possesses two metastable states: a strong-vortex regime and a weak-vortex regime. The present paper's chief goal is to address questions about the long-term climate statistics of rare events by way of a case study on SSW-like regime transitions in the Holton-Mass model: how often do they occur, what are their typical development pathways, and how variable are those pathways between events? 

We will use the framework of transition path theory \citep[TPT;][]{E2006towards}, which offers a concise set of quantities to answer these questions. An SSW event is represented as a \emph{transition path} from the strong vortex regime, which we denote state $A$, to the weak vortex regime, state $B$. The main quantity of interest will be the \emph{reactive current} $\bj_{AB}$, defined in section \ref{sec:current}, which specifies the flow of probability density through state space \emph{conditioned on an $A\to B$ transition event being underway}. To properly implement that conditional statement, we will need two auxiliary quantities. First, the \emph{forward committor} $\qp_B(\vx)$ gives the probability that the system, initialized in a state $\vx$, next reaches $B$ before $A$. This is a measure of progress toward SSW: what is the probability of observing a SSW before returning to the strong vortex climatology? Second, the \emph{backward committor} $\qm_A(\vx)$ gives the probability, looking backward in time, that the system visited $A$ more recently than $B$, i.e., the model was last in the meta-stable strong vortex climatology, as opposed to just recovering from a recent SSW. 

The forward committor itself was a primary focus of \cite{Finkel2021learning}, where we pursued forecasting as the main objective. Committor probabilities are generally gaining traction as a metric for weather prediction; see \cite{Tantet2015} for an application to atmospheric blocking, \cite{Lee2018environmentally} for an application to tropical cyclone downscaling, \cite{Lucente2022committor} for an application to El Ni{\~n}o,  and \cite{Miloshevich2022probabilistic} for an application to heat waves. However, in the present paper we are pursuing climatological statistics rather than forecasting probabilities, using the committor only as an intermediate calculation for the reactive current, which characterizes the full transition process from $A$ to $B$ rather than its ``forward half'' from $\vx$ to $B$. 

Some previous studies \citep{Crommelin2003,Tantet2015} have visualized what are essentially reactive currents for blocking events in an observable subspace of leading EOFs. However, these studies were not couched in the language of TPT, a formalism that brings more quantitative results. Namely, the reactive current $\bj_{AB}$ provides a direct estimate of the SSW rate, decomposing it over a continuous probability distribution of pathways. Formal TPT has not yet been widely taken up by the atmosphere-ocean science community, besides a few exceptions \citep{Finkel2020,Miron2020,Miron2022transition}. Part of our goal here is to encourage a common quantitative language for discussing regime transitions, which could help to organize several existing lines of research. 

$\bj_{AB}$, like $\qp_B$, can be expressed as a function of any observable subspace for visual exploration, with the complementary subspace treated as random variables. It is most enlightening to use observables with concrete physical meaning. A recent article \cite{Miloshevich2022probabilistic} exploited this property to interpret a neural-network-learned committor for heat waves in terms of geopotential height and soil moisture, thus quantifying their predictive power at various lead times. In \cite{Finkel2021learning}, we visualized the committor and expected lead time in a two-dimensional subspace consisting of zonal wind $U$, an index for polar vortex strength, and vertically integrated heat flux (IHF), which roughly measures the amplitude and phase tilt of vortex-disrupting planetary waves. Here we continue to use those coordinates, but also introduce a new subspace based on the zonal-mean meridional potential vorticity (PV) gradient and eddy enstrophy. These two quantities obey a conservation law in the absence of dissipation and stochastic forcing, a slight variation of the Eliassen-Palm relation. This allows us to diagnose more precisely the crucial roles of dissipation and stochastic forcing in driving the transition process, an important step toward understanding their causal relationship. Other kinds of atmospheric regime transitions will have different relevant physical diagnostics, any of which can be seen as an independent variable for the committor function and reactive current.

This paper is organized as follows. In section \ref{sec:model} we review the dynamical model. In section \ref{sec:current} we visualize the evolution of SSW events using the probability current, and introduce the key quantities for TPT---committors, densities, and currents---along with a brief summary of the method to compute them, which is more thoroughly explained in the supplementary document. In section \ref{sec:composite_U-IHF}, we use reactive current to construct a composite SSW evolution, and compare this to the standard composite method. In section \ref{sec:enstrophy_budget}, we change coordinates to better examine the dynamics of SSW events. We assess future directions and conclude in section \ref{sec:conclusion}.

\section{A stochastically forced Holton-Mass model of SSW dynamics}
\label{sec:model}

We use exactly the same model as in \cite{Finkel2021learning}, which is presented here for completeness. 

\subsection{Model specification}
\begin{figure*}
    \centering
    \includegraphics[trim={0cm 15cm 0cm 0cm}, clip,width=0.9\linewidth]{"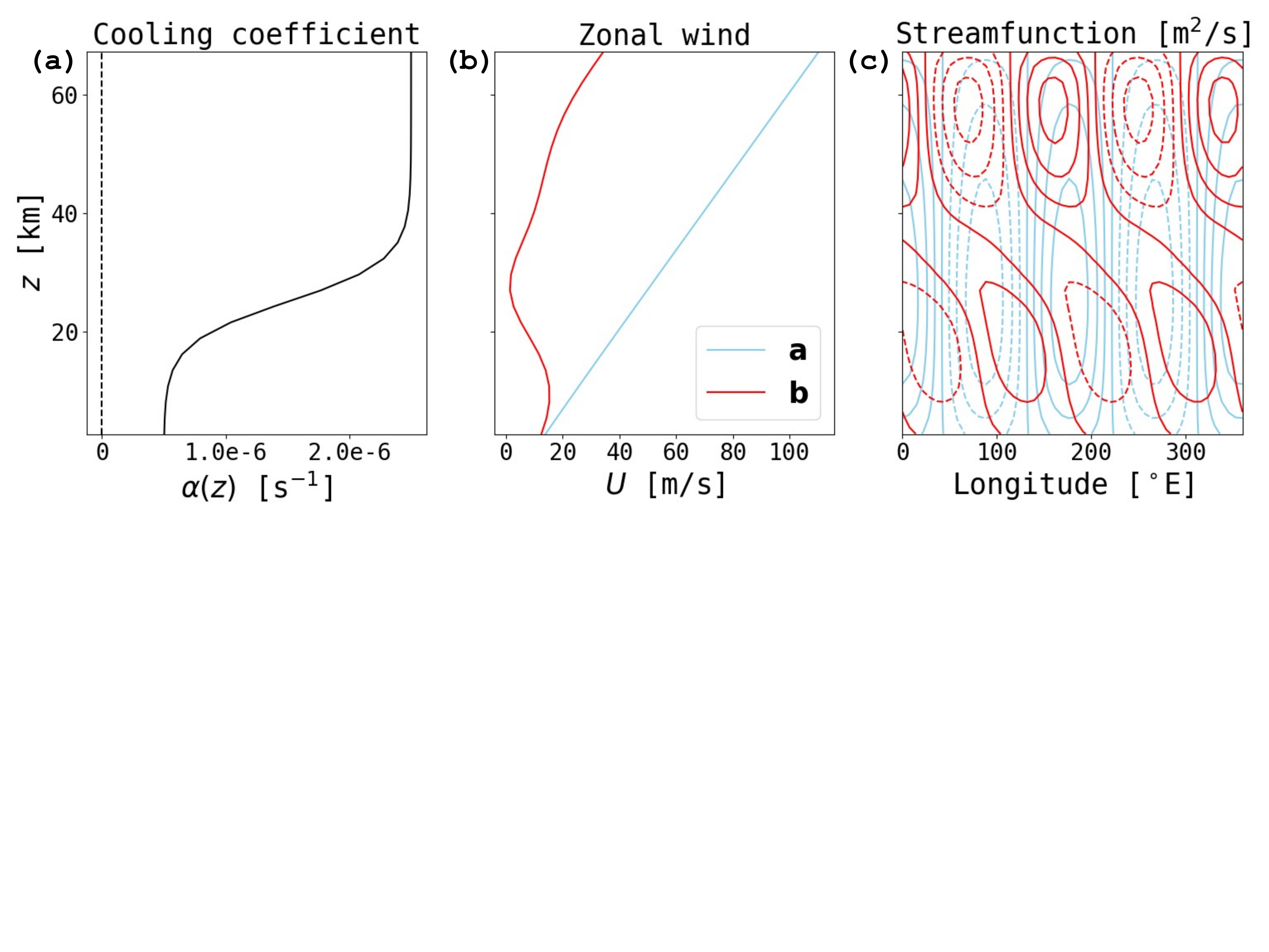"}
    \caption{\textbf{Parameters and stable equilibria of the Holton-Mass model}. (a) The Newtonian cooling profile $\alpha(z)$. (b) Zonal-mean zonal wind $U(z)$ and (c) perturbation streamfunction $\psi'(x,60^\circ\text{N},z)$, with contour spacing of $1.5\times10^7$ m$^2$/s. Dashed lines mean negative values. Blue indicates the strong vortex equilibrium, $\va$, and red indicates the weak vortex equilibrium, $\vb$, as in Eqs.~\eqref{eq:abdef}. }
    \label{fig:equilibria}
\end{figure*}

\cite{holton_mass} developed a minimal model for the variability of the winter stratospheric polar vortex, capturing the wave-mean flow interactions behind sudden stratospheric warming events. The model's prognostic variables consist of a zonally averaged zonal wind $\ov u(y,z,t)$ and a perturbation geostrophic streamfunction $\psi'(x,y,z,t)$ on a $\beta$-plane channel with a central latitude of $\theta=60^\circ$N, a meridional extent of 60$^\circ$, and a height of 70 km, with the coordinate $z$ ranging from 0 at the bottom of the domain (the tropopause) to 70 km at the top of the domain.  $\ov u$ and $\psi'$ are projected onto a single zonal wavenumber $k=2/(a\cos\theta)$ and a meridional wavenumber $\ell=3/a$:
\begin{align}
    \ov u(y,z,t)&=U(z,t)\sin(\ell y) \label{eq:meanflow_ansatz}\\
    \psi'(x,y,z,t)&=\re\{\Psi(z,t)e^{ikx}\}e^{z/2H}\sin(\ell y), \label{eq:wave_ansatz}
\end{align}
where $a=6370$ km is the radius of Earth, and $H=7$ km is the scale height. $U$ (the mean flow) and $\Psi$ (a complex-valued wave amplitude) evolve according to the projected primitive equations and the linearized quasi-geostrophic potential vorticity (QGPV) equation. A non-dimensionalized version of the equations is as follows, rearranged slightly from \cite{Finkel2021learning}. The mean flow $U(z,t)$ satisfies
\begin{subequations}
	\label{eq:wave_meanflow_eqns}
    \begin{align}
    	\frac{2}{(\eps\ell)^2}\partial_t&\Big[\gcal^2\beta+\eps\big(\gcal^2\ell^2U+U_z-U_{zz}\big)\Big] \label{eq:meanflow_eqn}\\
        &=\frac{2}{\eps\ell^2}e^z\partial_z\big[e^{-z}\alpha\partial_z(U-U^R)\big] \nonumber\\
        &\hspace{0.5cm}+ke^z\im\{\Psi^*\Psi_{zz}\}  \nonumber \\
        &\hspace{-1.5cm}\text{with boundary conditions} \nonumber\\
        U(z=0)&=U^R(z=0)=10\,\mathrm{ m/s} \nonumber\\
        U_z(z=z_{\mathrm{top}})&=U_z^R(z=z_{\mathrm{top}})=\gamma/1000 \nonumber
    \end{align}
    while the perturbation streamfunction amplitude $\Psi(z,t)$ satisfies
    \begin{align}
        (\partial_t+ik\eps U)&\bigg[-\gcal^2(k^2+\ell^2)-\frac14+\partial_z^2\bigg]\Psi \label{eq:wave_eqn}\\
        +ik\Psi&\Big[\gcal^2\beta+\eps\big(\gcal^2\ell^2U+U_z-U_{zz}\big)\Big] \nonumber\\
        &=-\bigg(\partial_z-\frac12\bigg)\bigg[\alpha\bigg(\partial_z+\frac12\bigg)\Psi\bigg]
        \nonumber\\
        &\hspace{-1.5cm}\text{with boundary conditions} \nonumber\\
        \Psi(z=0)&=\frac{gh}{f_0} \nonumber \\
        \Psi(z=z_{\mathrm{top}})&=0. \nonumber
    \end{align}
\end{subequations}
We have defined the nondimensional parameter $\gcal^2:=H^2N^2/(f_0^2L^2)$, where $f_0$ is the coriolis parameter at $60^\circ$N, $N^2=4\times10^{-4}$ is the the stratification, and $L=2.5\times10^5$ km is a horizontal length scale chosen to make non-dimensionalized $U$ and $\Psi$ variables have similar climatological variances. The linear relaxation towards $U^R(z)=10\,\mathrm{m/s}+(\gamma/1000)z$ on the right-hand side of Eq.~\eqref{eq:meanflow_eqn} is the force that maintains the typically strong polar vortex. Here $\gamma=1.5$ m s$\inv$ km$\inv$. The relaxation is mediated by a Newtonian cooling profile $\alpha(z)$, which is plotted in Fig. \ref{fig:equilibria}a, in its original dimensional units. Meanwhile, the lower boundary condition on $\Psi$ comes from a bottom topography $h\cos(kx)$, where $h=38.5$ m. This serves as a source of planetary waves. 

There are two differences from \cite{Finkel2021learning}, besides rearrangement. First, \cite{Finkel2021learning} had an erroneous but inconsequential negative sign in front of $U_{zz}^R$ (their Eq. 3) which is corrected in Eq.~\eqref{eq:meanflow_eqn}. Second, the left side of Eq.~\eqref{eq:wave_eqn} has two terms, $\pm ik\eps\gcal^2\ell^2U\Psi$, which could be cancelled out; we have retained them both to maintain a term-by-term correspondence with the original QGPV equation,
\begin{align}
	(\partial_t+\ov u\partial_x)q'+v'\partial_y\ov q&=\text{ sources $-$ sinks}, \label{eq:qgpv_pde} \\
	\text{where }q'&=\nabla^2\psi'+\fn e^{z/H}\partial_z(e^{-z/H}\psi')\\
	\text{and }v'&=\partial_x\psi'
\end{align}
which will be important when deriving the enstrophy budget in section \ref{sec:enstrophy_budget}.

After discretizing to 27 vertical levels, we end up with a state space of dimension $d=3\times(27-2)=75$, with a state vector
\begin{align}
    \bx(t)=\Big[\re\{\Psi(t)\},\im\{\Psi(t)\},U(t)\Big]\in\R^{75}
\end{align}
each of the three entries representing a vector with 25 discrete altitudes. We thus obtain a system of 75 ODEs, $\dot{\bx}(t)=\bm{v}(\bx(t))$. We furthermore perturb the system by stochastic forcing to represent unresolved processes such as smaller-scale Rossby and gravity waves, initial condition uncertainties, and sources of model error, an approach originally put forward by \cite{Birner2008} and used more recently by \cite{Esler2019}. The forcing is white in time, giving an It{\^o} diffusion
\begin{align}
	\label{eq:ito_diffusion}
	d\bx(t)=\bm v(\bx(t))\,dt + \bm\sigma(\bx(t))\,d\mathbf{W}(t)
\end{align}
where $\bm v(\vx)$ (not to be confused with meridional wind velocity, $v$) is the drift function determined by Eqs.~\eqref{eq:wave_meanflow_eqns}. $\mathbf{W}(t)$ is an $(m+1)$-dimensional white-noise process, and $\bm\sigma\in\R^{d\times(m+1)}$ is a matrix specifying the spatially smooth structure of the noise as Fourier modes in the vertical. $\bm\sigma$ could depend on the state vector $\bx$, but for simplicity we fix it to a constant, defined as follows. At each timestep $\delta t=0.005$ days, after incrementing the full system by $\delta\bx=\bm v(\bx)\delta t$, we additionally increment the zonal wind profile by
\begin{align}
	\label{eq:noise_structure}
	\delta U(z)=\sigma_U\sum_{k=0}^{m}\eta_k\sin\bigg[\bigg(k+\frac12\bigg)\pi\frac{z}{z_{\mathrm{top}}}\bigg]\sqrt{\delta t}
\end{align}
where $\sigma_U=1$ m s$\inv$ day$^{-1/2}$, whose units reflect the quadratic variation of Brownian motion. The numerical scheme is known as Euler-Maruyama \citep[see, e.g.,][ch. 5]{Pavliotis}. Equation \ref{eq:noise_structure} fully defines the matrix $\bm\sigma$. For $k=0,\hdots,m$, the $k$th column starts with 50 zeros, since there is no forcing on $\re\{\Psi\}$ or $\im\{\Psi\}$. The last 25 entries are evenly spaced samples of the sinusoidal factor in Eq.~\eqref{eq:noise_structure}, all times $\sigma_U$.

The specific choice of stochastic forcing does affect the transition path statistics, but our method can be applied to any stochastic forcing. Because of the nonlinear coupling between $U(z)$ and $\Psi(z)$ in Eqs.~\eqref{eq:meanflow_eqn} and~\eqref{eq:wave_eqn}, the noise injected to $U$ feeds to $\Psi$ after a single timestep.

\subsection{Diagnostics}

Until section \ref{sec:enstrophy_budget}, we use two main diagnostics for visualization, the same as in \cite{Finkel2021learning}. The first is zonal wind strength $U(z)$, an index for vortex strength which is used to define regimes $A$ and $B$. The second is the meridional eddy heat flux $\ov{v'T'}(z)$, which quantifies the heat being advected into the polar region associated with the sudden warming, and in the quasi-geostrophic limit, the vertical propagation of Rossby waves. In the Holton-Mass model, this takes the form
\begin{align}
	\ov{v'T'}(z)=\frac{Hf_0}{R}\ov{\pder{\psi'}{x}\pder{\psi'}{z}}\propto e^{z/H}|\Psi(z)|^2\pder{\varphi}{z},
\end{align}
where $R$ is the ideal gas constant for dry air and $\varphi$ is the phase of the complex-valued streamfunction $\Psi$. Hence the heat flux is related to the amplitude and phase tilt of the waves, both of which rise significantly during a SSW event. We also use the density-weighted vertical integral of heat flux, 
\begin{align}
	\text{IHF}(z):=\int_0^ze^{-z/H}\ov{v'T'}(z')\,dz'
\end{align}
which varies more smoothly than $\ov{v'T'}$ at any single altitude.

\subsection{Bistability}
We use the same constant parameters and boundary conditions as \cite{Finkel2021learning}, which give rise to two stable equilibria: a radiative equilibrium-like state, denoted $\va$, and a disturbed state $\vb$, in which upward propagating stationary waves flux momentum down to the lower boundary, weakening zonal winds. Detailed bifurcation analysis by \cite{Yoden1987_bif} and \cite{Christiansen2000} found a range of values for bottom topography $h$ that create bistability. Figure~\ref{fig:equilibria}(b,c) depicts the zonal wind and streamfunction of these two equilibria. SSW events in this model are abrupt transitions from the region near $\va$ to the region near $\vb$. If a strong wave from below happens to catch the stratospheric vortex in a vulnerable configuration, then a burst of wave activity can propagate upward, ripping apart the polar vortex and causing zonal wind to collapse \citep{Charney1961,Yoden1987_dyn}. With certain parameters, the vortex can get stuck in repeated ``vacillation cycles'', in which the vortex begins to restore with the help of radiative forcing, only to be undermined quickly by the wave. The situation of two well-separated equilibria is highly idealized, and not a generic feature of climate phenomena; this system, with these parameters, serves to demonstrate qualitative features of SSW, not represent the real stratosphere quantitatively. \cite{holton_mass,Yoden1987_dyn,Christiansen2000}, and \cite{Finkel2021learning} contain further details. 

A \emph{transition path} is defined as an unbroken segment, or trajectory, of the system that begins in a region $A$ of state space (a neighborhood of $\va$) and travels to another region $B$ (a neighborhood of $\vb$) without returning to $A$. As in \cite{Finkel2021learning}, we define $A$ and $B$ based on the zonal-mean zonal wind at $z=30$ km:
\begin{subequations}
    \label{eq:abdef}
    \begin{align}
        A&=\{\vx\in\R^d:U(30\text{ km})(\vx)\geq 53.8\text{ m/s}\} \label{eq:adef}\\
        B&=\{\vx\in\R^d:U(30 \text{ km})(\vx)\leq 1.75\text{ m/s}\} \label{eq:bdef}
    \end{align}
\end{subequations}
where the velocity thresholds correspond to the vortex strength at 30 km for the fixed points $\va$ and $\vb$, respectively.

An SSW event is then a transition from $A$ to $B$, while the reverse, from $B$ to $A$, represents the recovery of the vortex. The definition of $B$ modifies the widely used definition of \cite{cp07} in two ways. First, we use zonal wind at 30 km above the tropopause (in log-pressure coordinates), because 30 km is where the zonal wind profile of $\vb$ reaches a minimum; \cite{Christiansen2000} used this same coordinate when studying the same model. (The standard 10 hPa pressure level would correspond to $z=-7\,\mathrm{km}\times\log(10/1000)-10\,\mathrm{km}\approx22$ km above the troposphere in this model.) 
We also modify the zonal wind thresholds  order to ensure that $\va\in A$ and $\vb\in B$.

An important consequence of our $A$ and $B$ definitions is that the $A\to B$ transition path takes $\sim80$ days. By design, this includes the slow initial \emph{preconditioning} stage of vortex breakdown in advance of the $\sim10$-day time horizon that traditionally comprises an SSW event. In this paper, `SSW event' should be interpreted as both the preconditioning and the ensuing vortex collapse. 

\begin{figure*}
    \centering
    \includegraphics[trim={0cm 1.5cm 0cm 0cm}, clip,width=0.9\linewidth]{"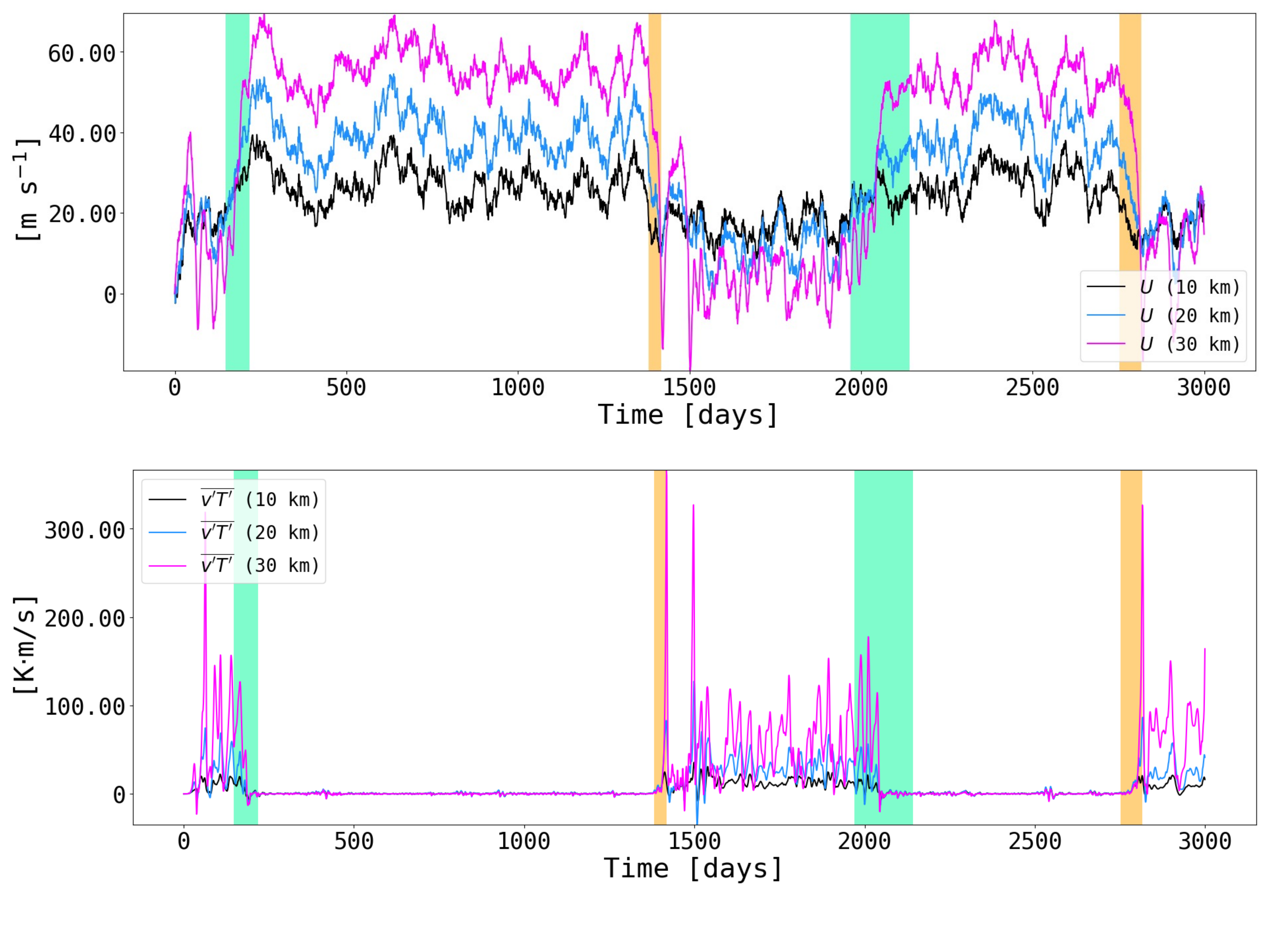"}
    \caption{\textbf{Regime transitions}. We plot (a) the zonal-wind strength $U$, and (b) the eddy heat flux $\ov{v'T'}$, over the first 3000 days of a long stochastic simulation. The quantities are evaluated at $z=10,20$, and 30 km. The time interval contains two transitions from $A$ (a strong vortex) to $B$ (a weak vortex) and back. $A\to B$ transitions are highlighted in orange, and $B\to A$ transitions are highlighted in green.}
    \label{fig:timeseries}
\end{figure*}

Figure \ref{fig:timeseries} shows timeseries of $U$ and $\ov{v'T'}$ at several different altitudes as the system goes through several transition paths in a long simulation. As in Fig. 2 of \cite{Finkel2021learning}, orange strips denote $A\to B$ transitions while green strips denote $B\to A$ transitions. The long periods in between, which we call the $A\to A$ and $B\to B$ phases, demonstrate the bistable nature of regimes $A$ and $B$. The fleeting $A\to B$ phase, however, is what we seek to understand. When the system is en route from $A$ to $B$, we say it is $(AB$)-\emph{reactive}, using a term from chemistry literature where the passage from $A$ (reactant) to $B$ (product) models a chemical reaction. The following section will introduce the \emph{reactive density} $\pi_{AB}(\vx)$ and associated \emph{reactive current} $\bj_{AB}(\vx)$ which help us visualize the transition as a path distribution through state space and make the foregoing observations more quantitative. 

\section{The reactive density and reactive current: A distribution over transition paths} 
\label{sec:current}

\begin{figure*}
    \centering
    \includegraphics[trim={0cm 5cm 0cm 0cm}, clip,width=0.99\linewidth]{"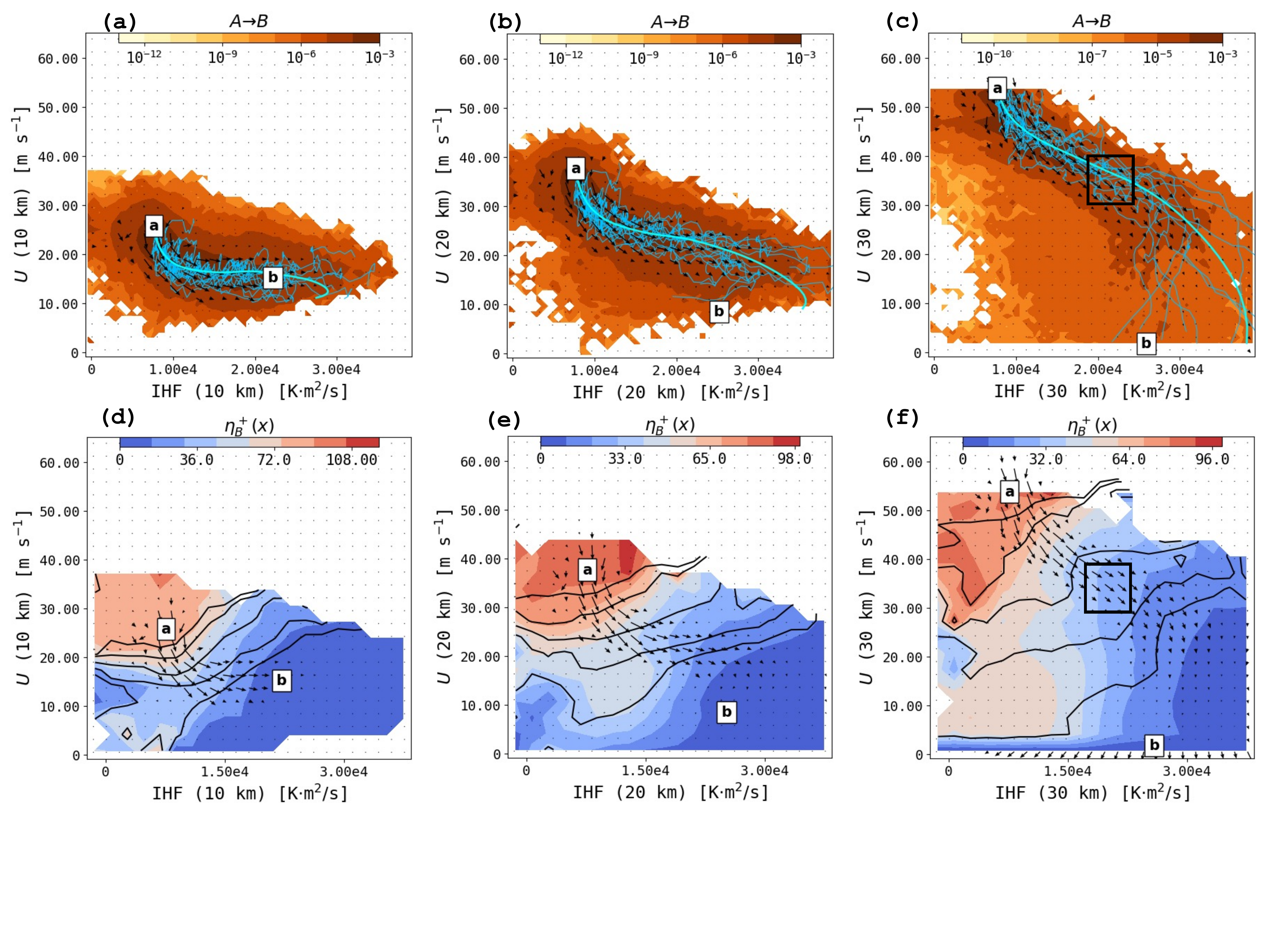"}
    \caption{\textbf{Currents, densities, committors, and expected lead times.} (a): Background shading is the reactive density $\pi_{AB}$, on a log scale. Thin blue lines are ten randomly selected transition paths from the long control simulation. Thick cyan curve is the minimum-action path from $A$ to $B$. Also overlaid is a vector field representing reactive current $\bj_{AB}$. The subspace is $(U,\ \text{IHF})$ evaluated at $z=10$ km. Positions of the fixed points $\va$ and $\vb$ are marked. Arrows represent $\bj_{AB}$. (b, c): Same as (a), but at $z=20$ and 30 km respectively. (d) The expected lead time $\tbd_B$ is shaded as background color, and level sets of the committor $\qp_B$ 0.1, 0.2, 0.5, 0.8, and 0.9 are overlaid as black curves. (e, f): Same as (d), but at $z=20$ km and 30 km respectively. A box marks a transition region between narrow, constrained current and wide, dispersed current. See text for a description.}
    \label{fig:projections_2d}
\end{figure*}

We consider the long-time behavior of our stochastic Holton-Mass model $\bx(t)$ undergoing transitions between states $A$ and $B$. Aggregating together statistics from only the transition paths yields a probability distribution called the \emph{reactive density} $\pi_{AB}(\vx)$, defined such that
\begin{align}
	\pi_{AB}(\vx)\,d\vx&=\pr\{\bx(t)\in d\vx|\bx(t)\text{ is in } \nonumber \\
	&\hspace{1cm}\text{transition from $A$ to $B$}\} 
\end{align}
where $d\vx$ is a small region about $\vx$. One could estimate $\pi_{AB}$ by binning samples from a long simulation, but including only those samples in transit directly from $A$ to $B$. Associated to $\pi_{AB}$ is a vector field called the \emph{reactive current} $\bj_{AB}(\vx)$, which quantifies the probability flux passing through $\vx$ per unit time only during transition paths. Roughly speaking, $\pi_{AB}$ specifies where transition paths go, and $\bj_{AB}$ specifies how they move. Below we define them formally, but Fig.~\ref{fig:projections_2d}(a-c) gives some intuition by projecting them on the subspace $(U,\text{IHF})$ at $z=$10, 20, and 30 km. Background shading indicates the strength of $\pi_{AB}$, and arrows indicate the magnitude and direction of $\bj_{AB}$. Overlaid in thin blue lines are ten randomly sampled transition paths from the long ergodic simulation. These sample paths cluster in the same regions of state space identified as high-probability under $\pi_{AB}$, and on average flow along the arrows, corroborating qualitatively that $\pi_{AB}(\vx)$ and $\bj_{AB}$ describe the location and evolution of the model in state space.  

The transition path ensemble shows marked differences between altitudes. At $z=$10 km, the vortex strength ($U$) of states $\va$ and $\vb$ is about the same, but the IHF is very distinct. The reactive current aligns with the IHF axis. Mathematically, this reflects the lower boundary condition $U(z=0)=U^R(z=0)$. Physically, this means that the heat flux due to the wave is the dominant physical process, with only small changes in zonal wind strength. The higher altitude of $z=30$ km, by contrast, exhibits a large reduction in zonal wind strength, but only in the late stages of the process. In fact, the pattern of reactive density $\pi_{AB}$ at $z=30$ km (panel c) tells us that this final deceleration is quite sudden: the magnitude of $\pi_{AB}$ is large near $A$, meaning transition paths linger there for a long time and only slowly crawl downward and to the right. But at the point IHF(30 km) $\approx2.5\times10^4$ K$\cdot$m/s, $U$(30 km) $\approx30$ m/s (the region marked by a dotted circle in panels c and f), $\pi_{AB}$ reduces in magnitude and the reactive current spreads out widely as it turns downward toward set $B$. This is a signal that the transition paths are becoming both faster and more variable. 

As a further point of comparison with $\bj_{AB}$, we have plotted the minimum-action pathway from $A$ to $B$ with thick cyan lines (section 3 of the supplement specifies the numerical method). This represents the most likely transition path in the low-noise limit \citep[e.g.,][]{Freidlin1970,E2004minimum,Forgoston2018}, and indeed it follows the direction of reactive current. With finite noise, however, the transition path ensemble spreads significantly around the minimum-action pathway, especially at the higher altitude of 30 km in the late stage of the transition process. Because of this, it is not possible for \emph{any} single pathway, mininimum-action or not, to meaningfully represent the full ensemble. 

We will show that the slow, initial phase of SSW involves \emph{preconditioning} of the vortex: gradual erosion of the wind field by the stochastic forcing into a configuration that is especially susceptible to wave propagation. Once the wave burst is triggered, it imparts swift changes to the entire zonal wind profile. However, the bulk of SSW progress, probabilistically speaking, occurs in the preconditioning phase. Below we make this qualitative description precise by relating the reactive current to the forecast functions from \cite{Finkel2021learning}: the committor and expected lead time metrics. 

\subsection{Mathematical relationship between current, committor, density, and rate}

To formalize the description above and interpret the current rigorously, some definitions are in order, including a brief recap of the quantities from \cite{Finkel2021learning}. Let us fix an initial condition $\bx(t_0)=\vx$ with a vortex that is neither strong nor fully broken down, so $\vx\notin A\cup B$. $\bx(t)$ will soon evolve into either $A$ or $B$, since both are attractive. The probability of hitting $B$ first is called the \emph{forward committor} (to $B$):
\begin{align}
    \qp_B(\vx)&=\pr_\vx\{\bx(\tau_{A\cup B}^+(t_0))\in B\} \label{eq:committor_definition_B} 
\end{align}
where the subscript $\vx$ denotes a conditional probability given $\bx(t_0)=\vx$, and $\tau_S^+(t_0)$ is the \emph{first hitting time} after $t_0$ to a set $S\subset\R^d$:
\begin{align}
    \tau_S^+(t_0)=\min\{t>t_0:\bx(t)\in S\}.
    \label{eq:first_hitting_time_definition}
\end{align}
Like the expected lead time introduced below, the committor (under various aliases) predates TPT as an object of interest in the study of rare events~\citep{du1998transition,bolhuis2002transition}. However, as we will see below, it is a key ingredient in any TPT analysis.

Our system is autonomous, with no external time-dependent forcing, so we can set $t_0=0$ and drop the argument from $\tau_{A\cup B}^+$ without loss of generality. The autonomous assumption can be relaxed, either by augmenting $\vx$ with a periodic variable for time (e.g., to include the seasonal cycle) or by augmenting $A$ and $B$ to include initial and terminal times (e.g., to better examine climate change effects). Periodic- and finite-time TPT has been presented formally in \cite{Helfmann2020}, and we have applied it to a dataset of state-of-the-art ensemble forecasts in \cite{Finkel2022revealing}. As a conceptual demonstration, however, the autonomous Holton-Mass model makes for a clearer exposition.

While $\tau_{A\cup B}^+$ itself is a random variable, one can take its expectation to obtain the \emph{expected lead time} (to $B$),
\begin{align}
	\tbd_B(\vx):=\ex_\vx[\tau_{A\cup B}^+|\tau_B^+<\tau_A^+],
\end{align} 
in other words, the expected time of arrival to $B$ conditional on hitting $B$ first. \cite{Finkel2021learning} described $\qp_B$ and $\tbd_B$ in detail, as they are central quantities for forecasting, and visualized them in their Figs. 2c,d and 3c in the observable subspace $(U, \text{IHF})$. We do the same here, but additionally we overlay the reactive current. In Fig.~\ref{fig:projections_2d}(d,e,f), background shading represents the expected lead time and black contours represent committor level sets of 0.1, 0.2, 0.5, 0.8, and 0.9.   

The committor's contour structure differs a lot between altitude levels. At 10 and 30 km (panels d and f), the contours have kinks. Depending on the initial condition, either a fluctuation in $U$ or IHF might have a greater effect on the committor. The intermediate altitude of 10 km seems special in having committor contours that align with the IHF axis along the main channel of reactive current. In other words, $\qp_B(\vx)$ is well-approximated by a linear function of $U$(20 km), which is consistent with the finding in \cite{Finkel2021learning} that the 21.5-km altitude holds the most predictive power for $\qp_B$.

$\bj_{AB}$ is related to $\qp_B$, generally flowing up the committor gradient. But $\bj_{AB}$ contains some key information that the committor does not. As a \emph{fore}cast function, the committor does not distinguish $A\to B$ transitions from $B\to B$ transitions, where the system leaves state $B$ (beginning to recover), but then falls back to the weak-vortex state. To isolate the transition events from $A$ to $B$, we need to introduce the \emph{backward committor} (to $A$):
\begin{align}
    \qm_A(\vx)=\pr_\vx\{\bx(\tau_{A\cup B}^-(t_0))\in A\}
\end{align}
where $\tau_S^-(t_0)$ is the \emph{most recent hitting time} 
\begin{align}
    \tau_S^-(t_0)=\max\{t<t_0:\bx(t)\in S\}
\end{align}
Intuitively, $\qm_A(\vx)$ is the probability of the system at point $\vx$ last came from $A$, not $B$. The backward-in-time probabilities refer specifically to the process $\bx(t)$ \emph{in steady-state}, allowing us once again to set $t_0=0$. In other words, $\qm_A(\vx)$ depends explicitly on the \emph{steady-state probability density} $\pi(\vx)$, where $\pi(\vx)\,d\vx=\pr\{\bx(t)\in\,d\vx\}$ is the long-term (climatological) probability of finding the system in a small region $d\vx$ about $\vx$. 

Having defined both forward and backward committors, we can express the reactive density as
\begin{align}
	\label{eq:reactive_density_definition}
	\pi_{AB}(\vx)=\frac{1}{Z_{AB}}\pi(\vx)\qm_A(\vx)\qp_B(\vx)  
\end{align}
where $Z_{AB}$ is a normalizing constant such that the right-hand side integrates to one. The associated reactive current can in turn be expressed
\begin{align}
	\label{eq:eqm_current}
	\bj_{AB}(\vx)&=\qm_A\qp_B\big[\pi\bm v-\nabla\cdot(\mathbf{D}\pi)\big] \\ 
	&\hspace{0.5cm}+\pi\mathbf{D}\big[\qm_A\nabla\qp_B-\qp_B\nabla\qm_A\big] \label{eq:reactive_current},
\end{align}
where the diffusion matrix $\mathbf{D}(\vx)=\frac12\bm\sigma(\vx)\bm\sigma(\vx)\tr$, and $\nabla$ represents the gradient operator over state space. 

Eq.~\eqref{eq:reactive_current} is a specific expression for the current of a diffusion process of the form~\eqref{eq:ito_diffusion}, which is the same general formulation as our model. But a more illuminating and general definition is its connection to the \emph{rate}, or inverse return time, of the event (approximately (1700 days)$\inv$ for the Holton-Mass model with our chosen parameters). Let $C$ be a closed hypersurface in $\R^d$ which encloses $A$ and is disjoint with $B$; we call this a \emph{dividing surface}. In the context of the diagrams in Fig.~\ref{fig:projections_2d}, $C$ is any curve separating region $A$ from region $B$. Then we have
\begin{align}
	\label{eq:jab_flux_density}
	\oint_C\bj_{AB}\cdot\vn\,dS=\text{Transition rate}
\end{align}
where $\vn$ is an outward unit normal from $C$ and $dS$ is a surface area element. The integral relationship~\eqref{eq:jab_flux_density} holds for any dividing surface, implying that the current is divergence-free outside of $A$ and $B$, but has a source in $A$ and a sink in $B$ (see \cite{VandenEijnden2006} for a thorough mathematical explanation of $\bj_{AB}$.) This constraint immediately implies a link between magnitude and width of $\bj_{AB}$ streamlines. In Fig.~\ref{fig:projections_2d}(c,f), the strong magnitude of $\bj_{AB}$ near $\va$ implies a thin central channel, and strict constraints on the mechanisms of early SSW onset. In other words, the initial preconditioning phase can only happen in a small number of ways. On the other hand, the subsequent weakening of $\bj_{AB}$ between $\qp_B=0.5$ and $\qp_B=0.8$ (in the boxed region of Fig.~\ref{fig:projections_2d}c,f) implies that paths fan out across state space, becoming more variable. This spreading, or diversity of events, is only with respect to $U$ and IHF at 30 km; at the lower altitudes, the current remains strong and narrow all the way through the transition process (Fig.~\ref{fig:projections_2d}, columns 1 and 2).

The reactive current and density characterize the transition path ensemble across the continuum of possible pathways, providing more information than the numerical value of the rate itself. Given any user-defined set of coordinates, the reactive current projection maps the transition paths in those coordinates, as a statistical ensemble with average behavior and variability. Below, following a brief note on the computational method, sections \ref{sec:composite_U-IHF} and \ref{sec:enstrophy_budget} demonstrate how to use reactive current and density to describe climatology and strengthen physical understanding of a rare transition event. 

\subsection{Computational method}
The quantities presented in section \ref{sec:current}, as well as the results to follow, could be computed directly by running a model for long enough to undergo a large number of SSW events and analyzing the statistics of those transitions. This procedure, which we call the ``ergodic simulation'' (ES) method, is possible in the 75-dimensional Holton-Mass model, and we have performed such a simulation of $10^6$ days for validation purposes. However, this can be a major computational barrier in global climate models when the numerical integration is costly and the return period is long compared to the simulation timestep. Anticipating the need for fundamentally different techniques in high-dimensional state spaces, we have instead used the Dynamical Galerkin Approximation \citep[DGA;][]{dga,Strahan2020}. A large collection of trajectories are launched in parallel with initial conditions distributed across state space, each one running for only a short time relative to the return period. Here we use $3\times10^5$ trajectories of length 20 days each, which is shorter than the 80-day duration of a single SSW event and much shorter than the 1700-day return period. Afterward, we assemble all these pieces together to estimate the quantities of interest, exploiting the Markov property. The total simulation time is not always reduced by this method---in our case, the short simulations total $6\times10^6$ days compared with the $1\times10^6$-day ES---but the format opens the door for many interesting possibilities, such as massive parallelization and adaptive sampling. In particular, as we show in \cite{Finkel2022revealing}, DGA is uniquely positioned to exploit large ensembles of short weather forecasts from high-fidelity operational models.

The basic DGA algorithm for rare event analysis has been  described and tested in a recent series of articles \citep{dga,Strahan2020,Finkel2021learning, antoszewski2021Phenol}.
It is closely related to the ``analogue Markov chain'' approach of \cite{lucente2021coupling}. 
Recently, an approach to learning neural network approximations of forecast functions using short trajectory data was introduced in~\cite{Strahan2022nnforecast}.
Due to the dependence on steady state and backward-in-time quantities, a full TPT analysis as carried out in this paper requires additional calculations beyond what is described in \cite{Finkel2021learning}.  We leave these details to the supplement in order to keep the focus on the results of our TPT analysis, which are robust with respect to algorithmic parameters.

\section{SSW composites}
\label{sec:composite_U-IHF}

Here we explain the traditional notion of a rare event `composite' and contrast it with the composite intrinsically defined by TPT. The results are qualitatively similar, but the TPT description allows a rigorous mathematical connection to the reactive current and SSW rate. 

\begin{figure}
    \centering
    \includegraphics[trim={0cm 0cm 33cm 0cm}, clip,width=0.8\linewidth]{"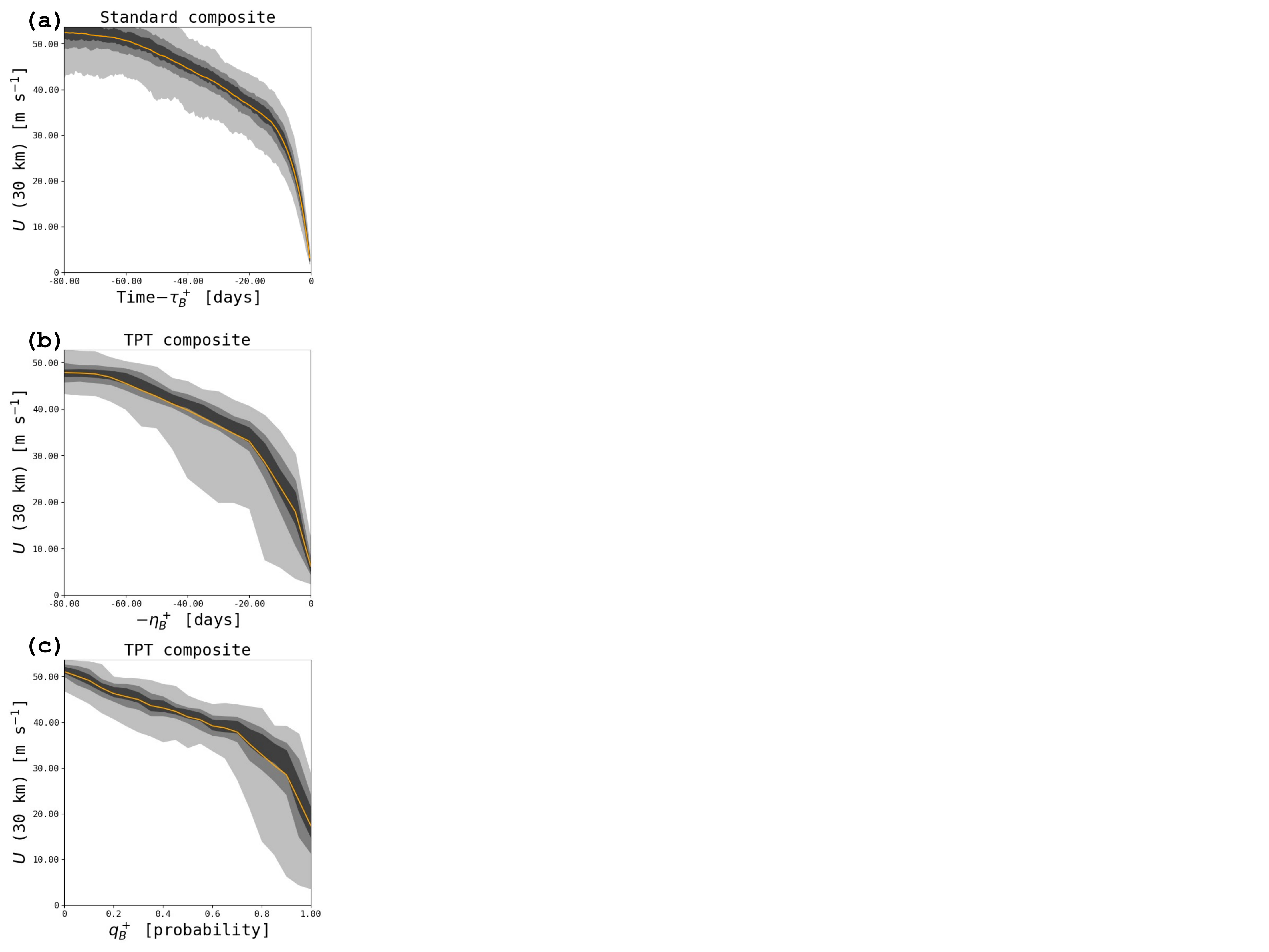"}
    \caption{\textbf{Composites evolution of SSW events}. Orange curves plot the mean value of $U$(30 km) at a given stage in the transition process; expanding gray envelopes show the middle 25-, 50-, and 90-percentile ranges. We use three different notions of progress: hitting time to $B$ ($t-\tau_B^+$, panel a), expected hitting time to $B$ ($-\eta_B$, panel b), and committor ($\qp_B$, panel c).}
    \label{fig:composites}    
\end{figure}

The standard ``composite'' of an SSW event is a day-by-day aggregate of all the SSW events in  a given dataset, aligned by the central warming date. This can include statistics, such as the mean and quantiles, of any observable function, such as the zonal-mean zonal wind or heat flux. \cite{cp07} and \cite{charlton_polvani_part2} used this method to describe SSW climatology and establish benchmarks for stratosphere-resolving GCMs. We form a standard composite of $U$(30 km) from our Holton-Mass model in Fig.~\ref{fig:composites}a, averaging together 300 events from a long ergodic simulation.

Here, we propose a complementary ``TPT composite'' based on reactive density. Instead of aligning events by the central warming date, we align the events by a general coordinate $f(\vx)$, which can be user-defined but must fulfill the minimal criterion of increasing from $A$ to $B$, so it represents some objective notion of progress. At any progress level $f_0$, the TPT composite is defined by restricting the reactive density $\pi_{AB}(\vx)$ to the level set $\{\vx:f(\vx)=f_0\}$. Fixing $f=f_0$ is not the same as fixing the lead time $\tau_B^+$, because the threshold might be crossed at different times by different transition paths. Note that $f(\vx)$ is a deterministic function of initial condition $\vx$, unlike the hitting time $\tau_B^+$, which is a random variable that changes between realizations launched from the same initial condition. Therefore, $\tau_B^+$ cannot itself be used as a progress coordinate.

In Fig.~\ref{fig:composites}b,c, we juxtapose alternative composites with the standard warming date coordinate $-\tau_B^+$. In panel b, we aggregate paths based on the negative expected lead time $-\tbd_B$ defined above: the \emph{expected} time until the central warming date. $-\tbd_B$ is the deterministic progress function that is closest (in the mean-square sense) to the random progress function $t-\tau_B^+$ defining traditional composites. Panel c uses an altogether different progress metric, the committor $\qp_B$ itself, which increases from 0 on $A$ to 1 on $B$.

The traditional and TPT composites are similar in shape, with an initially gradual decay in $U$(30 km) accelerating into a rapid decline in the final few days. As a function of $-\tbd_B$,  $U$(30 km) accelerates steadily through the whole transition, in both the traditional and TPT composites. But as a function of committor, $U$(30 km) decreases linearly at first and then accelerates downward between $\qp_B=0.6$ and $\qp_B=0.7$. According to the standard composite, $U$(30 km) becomes steadily less variable over time, with the whole ensemble collapsing into a single path by construction, as $t=0$ is the time of the event when $U$(30 km)$=0$. But when viewed as a function of expected lead time or committor, $U$(30 km) becomes more variable in the middle of the path, starting at $\tbd_B\approx50$ days or $\qp_B\approx0.65$ and lasting until the end, when $\tbd_B\to0$ and $\qp\to1$. 

The same variability is reflected in Fig.~\ref{fig:projections_2d}c,f. In the boxed region, the reactive density weakens and the reactive current spreads out, some paths turning straight downward into $B$ and others accumulating still more heat flux before making the plunge. The $\qp_B$ and $\tbd_B$ contours in Fig.~\ref{fig:projections_2d}f convey geometrically how it is possible to have such wide variation in zonal wind strength even at a fixed expected lead time. Along the central channel of strong reactive current, where most of the transition paths flow, the committor and expected lead time have an approximately (negative) linear relationship. But in the weak-$U$ flank of the current, especially in the boxed region, the $\qp_B$ level sets ``unkink'' to align with the IHF axis while the $\tbd_B$ level sets turn downward to align with the $U$ axis. The lowest visible level set of $\tbd_B$ thus spans a range of vortex strengths of $U$(30 km). 

Physically, the TPT composites are more variable than the traditional composite because $-\tbd_B$, the expected lead time---a deterministic function---is a coarser description than $t-\tau_B^+$, a random variable. The former is an average over all realizations, while the latter takes on a specific value for each realization, which is not actually known until after the warming occurs. Given only information on the resolved variables $\Psi(z,t)$ and $U(z,t)$ at a given time, the TPT composite is the best one can do. The expected lead time quantifies SSW predictability, as established in \cite{Finkel2021learning}. Here, we additionally incorporate the backward committor $\qm_A$ via the reactive density $\pi_{AB}$, and so restrict focus to \emph{transition} events---``major warmings''---from $A$ to $B$.

As a loose analogy, a student's progress toward a degree can be measured objectively in course credits. On the other hand, first-year exams might weed out half of all students, which means that the \emph{probabilistic} half-way point usually comes before half of required credits are done. A third metric, the time until graduation, can vary due to random effects like gap years and pandemics, which can cause a student to space their course load unevenly in time. Each cross-section of the student population---conditioning on a fixed number of credits completed, probability of graduation, or expected time until graduation---is a different statistical ensemble, each one conveying different information. 

Going forward, we will use the committor as the progress coordinate of choice. That way, each point along the composite is an average over trajectories that are equally predictable in their probability to reach $B$, i.e., to proceed to an SSW. Often it is not just a singular coin toss that determines the fate of $\bx(t)$, but a whole sequence of `coin tosses'---random turns through state space---aligning in just such a way to navigate from $A$ to $B$. With the committor as a progress coordinate, the `coin tosses' are equidistributed along the horizontal axis, though they may not be equidistributed in time. 

\begin{figure*}
    \centering
    \includegraphics[trim={0cm 0cm 25cm 0cm}, clip,width=0.65\linewidth]{"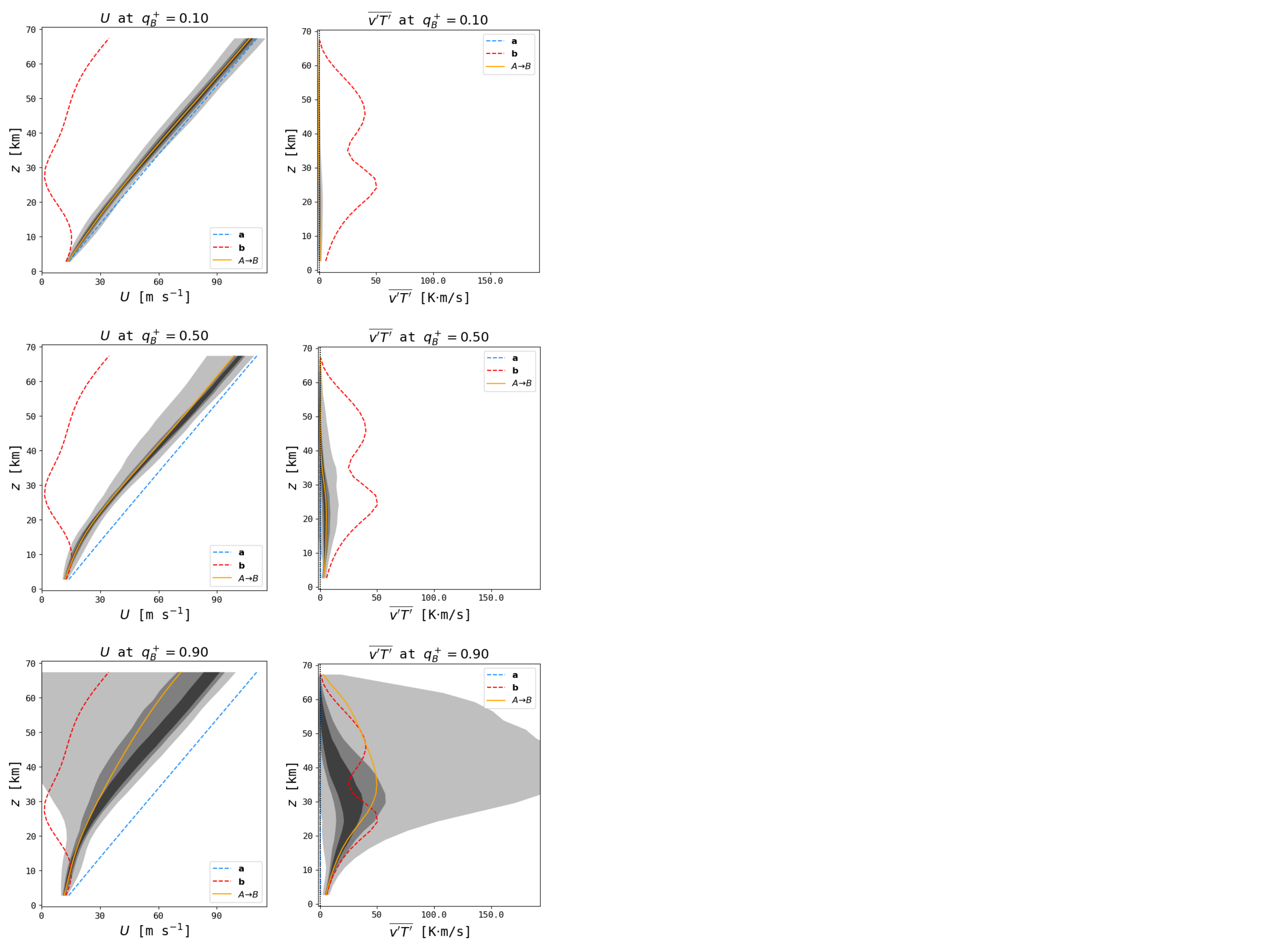"}
    \caption{\textbf{Vertical profiles of transition states and tendencies.} Left column: $U(z)$ averaged over $\qp_B=0.1$, $0.5$, and $0.9$. Orange curve is the mean, and gray envelopes represent the middle 25-, 50-, and 90-percentile ranges. Dashed blue and red curves represent $U(z)$ for the fixed points $\va$ and $\vb$. Right column: same as left, but for eddy meridional heat flux $\ov{v'T'}$. }
    \label{fig:profile_distributions}
\end{figure*}

The same composite technique can be used to visualize the vertical wind structure at different stages. Fig.~\ref{fig:profile_distributions} plots $U(z)$ and $\ov{v'T'}(z)$ as altitude-indexed probability distributions at a series of committor level sets: $\qp_B=0.1$, 0.5, and 0.9. The widening variability with increasing committor is faintly visible at low altitudes, but increases dramatically above $\sim23$ km, where at the $\qp_B=0.9$ level, the mean state (orange curve) falls well below the median state (central gray envelope.) This means the distribution of transition states is skewed left by a minority of paths with early collapse of upper-level winds. At the same committor range of 0.5-0.9, the vertical profile of meridional heat flux inflates dramatically. The altitude range of $z=$ 20-25 km is the key transition region, below which zonal wind evolves relatively smoothly and with a symmetric distribution, and above which it varies rapidly with a skewed distribution. $\ov{v'T'}(z)$ is maximum near this altitude. We speculate that the underlying reason is the Newtonian cooling profile $\alpha(z)$, which has its own transition region centered at 25 km. It is not surprising that zonal wind just below, at 21.5 km, is an optimal linear predictor, as we found in \cite{Finkel2021learning}.  

\section{A wave-mean flow interaction perspective}
\label{sec:enstrophy_budget}

The previous section presented $\bj_{AB}$ and $\pi_{AB}$ as functions of two basic observables, zonal wind and integrated heat flux, and constructed a composite evolution of these observables. In this section, we incorporate more detailed physical knowledge to improve the interpretability of our TPT results. In particular, we manipulate the the dynamical equations to derive an enstrophy budget in the Holton-Mass model, which reveals a more natural set of coordinates that separates conservative from non-conservative processes. By visualizing the current in these coordinates, we identify physical drivers of each stage in the transition process. Our goal is twofold: first, to show how TPT can be formulated for any observables, and second, more narrowly in the context of this study, how the dynamics become more clear when those observables are well-chosen.

\subsection{An eddy enstrophy formulation of the Holton-Mass model}

A common diagnostic for wave-mean flow interaction systems is the wave activity, $\acal=\rho_s\ov{q'^2}/(2\partial_y\ov q)$, whose evolution is related to the Eliassen-Palm (EP) flux divergence \citep{Andrews1976planetary}. \cite{Yoden1987_dyn} used wave activity extensively to analyze the vacillating regime (our set $B$) of the Holton-Mass model, in particular the upward wave propagation that destabilizes the vortex. Below we derive a related set of equations for the eddy enstrophy, which enjoys a simpler balance equation and which we have found is better numerically suited for TPT analysis.

The first step in deriving the EP relation is to multiply the QGPV equation~\eqref{eq:qgpv_pde} by $q'$ and take a zonal average, yielding
\begin{align}
	\partial_t\bigg(\frac{\ov{q'^2}}{2}\bigg)+\ov{v'q'}\partial_y\ov q&=\ov{q'(\text{sources}-\text{sinks})}
	\label{eq:enstrophy_budget_continuous}
\end{align}

We wish to work with the projected version of the equation, Eq.~\eqref{eq:wave_eqn}, rather than the original PDE, to account for the approximation $\sin^2(\ell y)\approx\eps\sin(\ell y)$ introduced by \cite{holton_mass} for projecting quadratic nonlinearities. The procedure is summarized below, and spelled out more thoroughly in section 4 of the supplement. 

Because of the ansatz~\eqref{eq:wave_ansatz}, $q'$ is represented in the projected equations by
\begin{align}
	q'\longleftrightarrow&\bigg[-\gcal^2(k^2+\ell^2)-\frac14+\partial_z^2\bigg]\Psi \label{eq:pvproj} \\
	&=:(-\delta+\partial_z^2)\Psi \nonumber
\end{align}
where $\longleftrightarrow$ denotes correspondence between the full governing equations and the projected, non-dimensionalized equations in the Holton-Mass model. Recall that $\Psi$ is the complex amplitude for the zonal-perturbation streamfunction $\psi'(x,y,z,t)$, in geostrophic balance with the wind $(u,v)$. 

As a general rule, the zonal average of the product of two wave quantities $\psi'_1$ and $\psi'_2$ of the form in Eq.~\eqref{eq:wave_ansatz}.is found by the following formula: 
\begin{align}
    \label{eq:zonal_correlation_rule}
	\ov{\psi_1'\psi_2'}&=\ov{\re\{\Psi_1e^{ikx}\}}\ov{\re\{\Psi_2e^{ikx}\}} \\
	&=\re\{\Psi_1^*\Psi_2\} \nonumber
\end{align}
Therefore, we multiply both sides of Eq.~\eqref{eq:wave_eqn} by the complex conjugate of~\eqref{eq:pvproj} and take the real part to obtain

\begin{subequations}
\label{eq:enstrophy_budget} 
    \begin{align}
    	&\partial_t\ecal+F_q\beta_e=D 
    \end{align}
    where
    \begin{align}
    	\ecal&=\frac12e^z\big|\big(-\delta+\partial_z^2\big)\Psi\big|^2 \\
    	&\longleftrightarrow \frac12\ov{q'^2} \nonumber
    \end{align}
    represents the eddy enstrophy; 
    \begin{align}
    	F_q&=ke^z\im\{\Psi^*\Psi_{zz}\} \\
    	&\longleftrightarrow\ov{v'q'} \nonumber
    \end{align}
    represents the meridional eddy PV flux;
    \begin{align}
    	\beta_e&=\gcal^2\beta+\eps\big(\gcal^2\ell^2U+U_z-U_{zz}\big) \\
    	&\longleftrightarrow \partial_y\ov q \nonumber
    \end{align}
    represents the meridional PV gradient; and
    \begin{align}
        D&=-\re\bigg\{e^z\big[\big(-\delta+\partial_z^2\big)\Psi^*\big]\times \nonumber\\
        &\hspace{1.0cm}\bigg(\partial_z-\frac12\bigg)\bigg[\alpha\bigg(\partial_z+\frac12\bigg)\Psi\bigg]\bigg\} \nonumber\\
        &\longleftrightarrow \ov{q'(\text{sources $-$ sinks})} \nonumber 
    \end{align}
    represents the production and dissipation of enstrophy.
\end{subequations}

The standard EP relation would be found by dividing both sides by the meridional PV gradient $\beta_e$, as in \cite{Yoden1987_dyn}. Instead, we next turn to the mean-flow equation~\eqref{eq:meanflow_eqn}, which is an evolution equation for the PV gradient $\beta_e$ rather than $U$ directly. Multiplying through by $\beta_e$, we find
\begin{subequations}
\label{eq:gramps_budget}
    \begin{align}
    	\partial_t\gramps&=R\beta_e+F_q\beta_e 
    \end{align}
    where
    \begin{align}
    	\gramps&:=\bigg(\frac{\beta_e}{\eps\ell}\bigg)^2 \\
    	R&:=\frac{2}{\eps\ell^2}e^z\partial_z\big[e^{-z}\alpha\partial_z(U-U_R)\big]
    \end{align}
\end{subequations}
The new quantity $\gramps(z)$ is the squared meridional gradient of zonal-mean potential vorticity, which is highly correlated to zonal wind strength $U(z)$ in the Holton-Mass model. $R$ is a relaxation coefficient for $\gramps$, strengthening the vortex via radiative cooling.

The advantage of this alternative EP relation is now clear: adding together Eqs.~\eqref{eq:enstrophy_budget} and~\eqref{eq:gramps_budget}, the meridional PV transport $F_q\beta_e$ cancels to give
\begin{align}
	\partial_t(\gramps+\ecal)&=R\beta_e+D. \label{eq:gebudget}
\end{align}
In this form, all the dissipative effects are contained on the right-hand side via the cooling coefficient $\alpha(z)$, which appears both in $D$ and $R$. $\gramps+\ecal$ would conserved, at every altitude separately, in the absence of dissipation and stochastic forcing. In this limit, an increase in eddy enstrophy $\ecal$ can only occur at the expense of the mean PV gradient characterized by $\gramps$. Of course, both non-conservative effects---dissipation and stochastic forcing---are critically important; vacillation cycles and transitions are possible only because the Holton-Mass model, like the full atmosphere, is an open system. The utility of Eq.~\eqref{eq:gebudget} is to isolate those nonconservative effects as almost extrinsic inputs. 


\subsection{Using the reactive current to quantify the importance of non-conservative processes}

\begin{figure*}
    \centering
    \includegraphics[trim={0cm 4cm 0cm 0cm}, clip,width=0.95\linewidth]{"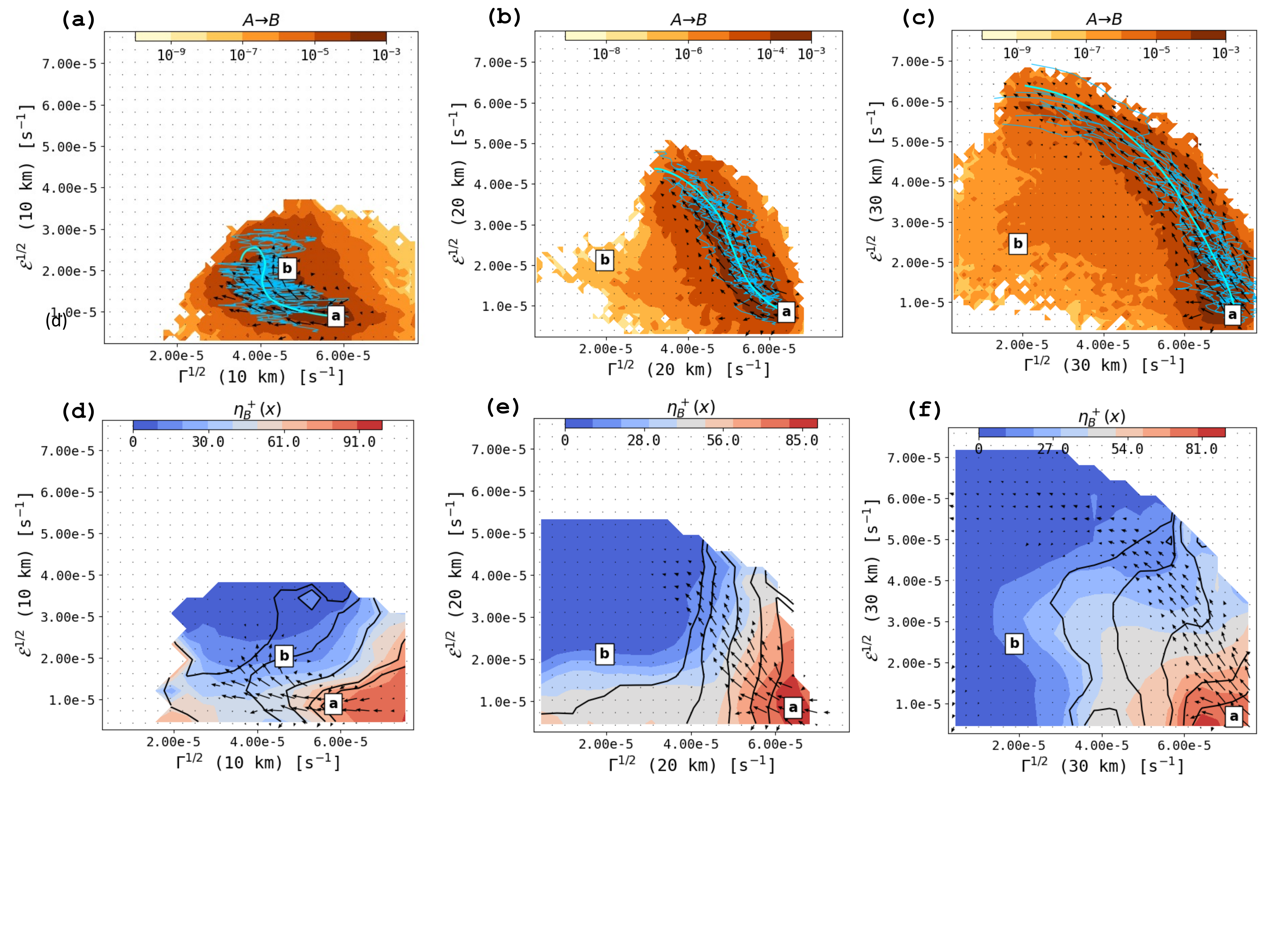"}
    \caption{\textbf{Current in wave-mean flow coordinates.} Same as Fig.~\ref{fig:projections_2d}, but for a different observable subspace ($\gramps^{1/2},\ecal^{1/2}$) instead of ($U$, IHF). See text for definitions. Eddies are characterized by RMS perturbation PV, $\ecal^{1/2}$, and the mean flow by the zonal mean PV gradient, $\gramps^{1/2}$. }
    \label{fig:current_gebudget}
\end{figure*}

Dissipation and forcing act to disrupt the conservation of $\gramps+\ecal$, with a specific pattern shown in Fig.~\ref{fig:current_gebudget}. The reactive current is shown at three altitudes, as in Fig.~\ref{fig:projections_2d}, but this time in the space $(\gramps^{1/2},\ecal^{1/2})$ instead of ($U$, IHF). We take square roots because the visualizations are more clear, and the units of s$\inv$ are more comparable with those of zonal wind $U(z)$ and radiative cooling $\alpha(z)$. (We note that the fixed point $\vb$ in panel (d) appears to have committor $<1$; this is possible when projecting out nonlinear coordinates because set $B$ is defined based on the 30-km level, and the state-space regions that resemble $\vb$ at 10 km may not resemble it at 30 km.) In the upper stratosphere, at $z=30$ km (panels c and f), the main channel of reactive current flows along a circular arc, approximately conserving $\gramps+\ecal$, all the way through the $\qp_B=0.9$ surface: the evolution of an SSW is a nearly conservative interaction between waves and the mean flow right up to the end. Then, the current weakens in magnitude and spreads out, indicating the critical non-conservative processes at the end, where the breaking and dissipation of the anomalous waves cements the SSW event. Just as in the ($U$,IHF) space, the reactive density $\pi_{AB}$ decreases along that circular arc, meaning the transition paths accelerate. 

On the other hand, $\bj_{AB}$ projected at $z=10$ km (panels a and d) shows that the dynamics are never conservative in the lower stratosphere: the initial motion points not along a circular arc but directly leftward, such that $\gramps+\ecal$ is decreasing from the start. From the enstrophy budget~\eqref{eq:gebudget}, we conclude that a combination of dissipation and stochastic forcing acts strongly at 10 km to precondition the vortex. The next subsection shows that stochastic forcing plays the more decisive role. 

Finally, consider the middle altitude of 20 km, where $\bj_{AB}$ has a shape that is intermediate between the current at 10 and 30 km. It does not have distinctly positive or negative curvature, but flows along a straight channel from $A$ to $B$. 20 km seems to be in just the right altitude range to feel significant dissipation and stochastic forcing---a feature of the lower boundary---but also to channel a good share of the loss of $\gramps$ to the gain of $\ecal$, a quasi-conservative property of the loftier 30 km. The resulting committor, expected lead time, and reactive current are approximately linear functions of $\gramps^{1/2}$(20 km) and $\ecal^{1/2}$(20 km). Indeed, the wind and heat flux at 20 km were the most useful for prediction in \citep[][their section 4]{Finkel2021learning}.

\begin{figure*}
    \centering
    \includegraphics[trim={0cm 3cm 0cm 0}, clip,width=0.9\linewidth]{"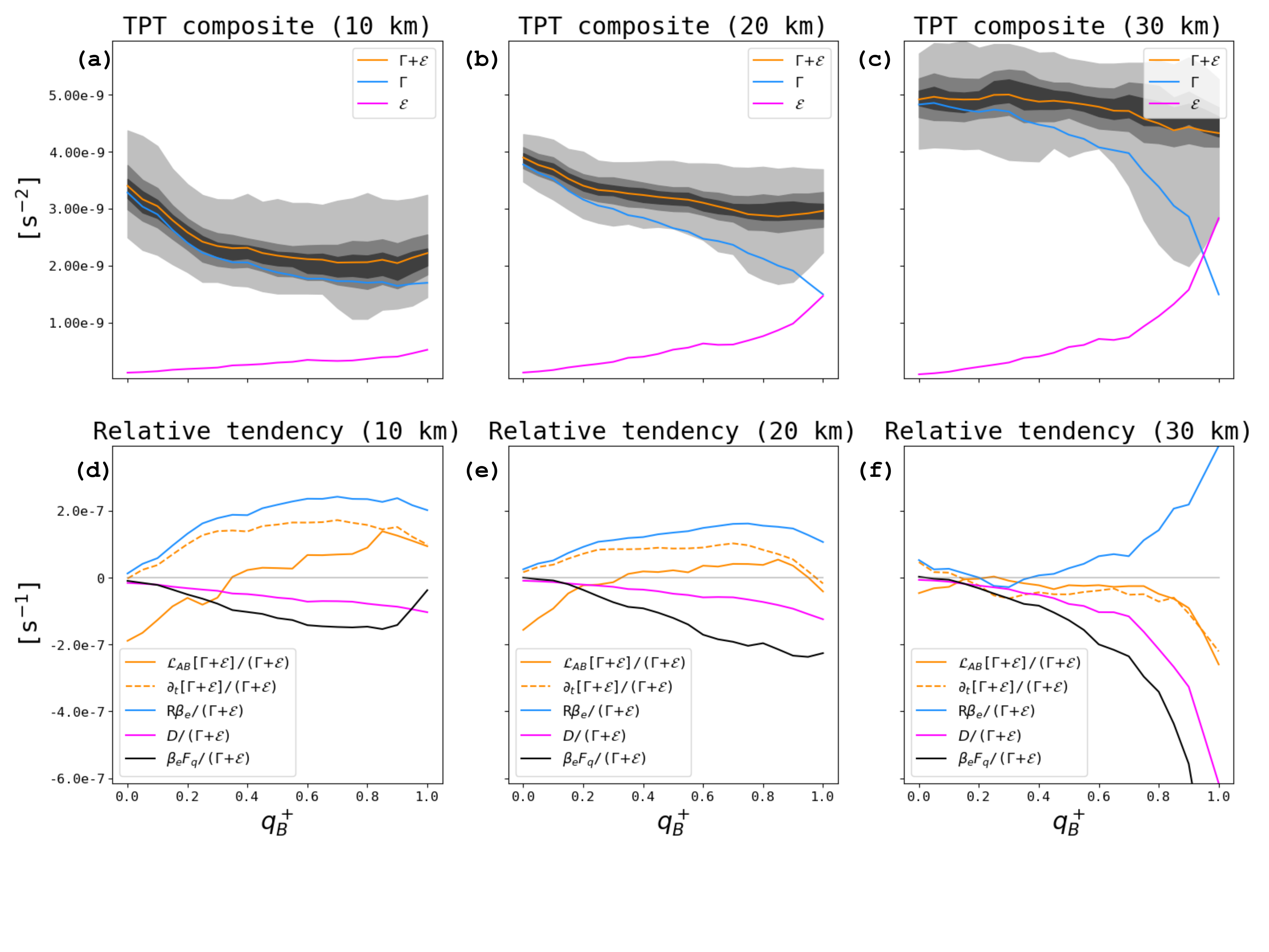"}
    \caption{\textbf{Enstrophy budget analysis through the $A\to B$ transition}. (a) Blue, pink, and orange curves represent mean values of $\gramps$, $\ecal$, and their sum at $z=10$ km, conditioned on the system being in a transition path and near a given committor level (which varies along the horizontal axis). Gray envelopes represent the middle 25, 50, and 90-percentile ranges of $\gramps+\ecal$; when the orange curve is not at the center of the gray envelopes, the distribution is skewed. (b, c): same as (a), but at $z=20$ and 30 km respectively. (d) Solid orange curve shows the expected tendency of $\gramps+\ecal$ at 10 km, again conditioned on being in a transition path and near a given committor level. Dashed orange curve shows the deterministic tendency at the same committor levels; the difference between the two indicates the role of stochastic forcing. Blue curve shows the relaxation of $\gramps$ (the squared meridional PV gradient), pink curve shows the dissipation of enstrophy, and black curve shows the meridional transport of PV, $F_q\beta_e$, which when negative indicates a gain for $\ecal$ at the expense of $\gramps$. The sum of the blue and pink curves gives the dashed orange curve. (e, f): same as (d), but at $z=10$ and 20 km respectively. All tendencies are normalized by $\gramps+\ecal$, as the legend shows, for a comparable vertical scale across altitudes.}
    \label{fig:gebudget_analysis}
\end{figure*}

Fig.~\ref{fig:gebudget_analysis}a,b,c show the composite evolution of $\gramps+\ecal$ in orange, along with $\gramps$ in blue and $\ecal$ in pink, at the same three altitudes 10, 20, and 30 km. All three altitudes show evidence of dissipation, with $\gramps+\ecal$ weakening as the committor increases, but with distinct differences in the rates. The $\gramps+\ecal$ composite is concave up at 10 km, implying dissipation is most important at the early stage, when the predictability of the event is limited. At 20 km, the composite is practically linear, implying that dissipation maintains a constant role in the event's evolution, gradually opening the valve to wave propagation at the last stage of the event. At 30 km, the composite is concave down: the flow is initially conservative, with exchange between mean flow and eddies at the onset of vortex breakdown, followed by strong dissipation of the waves when the event is all but assured.

At 20 and 30 km, the distribution of $\gramps+\ecal$ begins symmetric, with the mean (orange) tracking the median (near the center of the dark gray band). Then between $\qp_B=0.6$ and 0.7, the lower tail of the distribution expands quickly, skewing the distribution negative. The distribution at 10 km maintains a slight negative skew for the entire transition path. The skewness reflects the occurrence of ``minor warmings'' preceding the SSW, when the vortex begins to break down, but partially recovers before the final event. 

The composites, as well as the reactive currents, support the notion of the ``typical'' transition path as an initially non-conservative creep at low altitudes, opening up a valve to allow waves to propagate upward, finally yielding a very abrupt collapse at high altitudes follows after a long, mostly conservative phase. With the enstrophy budget~\eqref{eq:gebudget}, we can assess the importance of each term by plotting those composites as well.  Fig.~\ref{fig:gebudget_analysis}d,e,f show the composite evolution of each term at each altitude: $R\beta_e$ (the relaxation of the squared mean PV gradient, $\gramps$) in blue, $D$ (the dissipation of enstrophy, $\ecal$) in pink, and $\beta_eF_q$ (the transfer of enstrophy from $\gramps$ to $\ecal$) in black, all normalized by the total $\gramps+\ecal$ at each level to account for the altitude-dependent differences in variability. This allows us to compare how strong each dissipative force is \emph{relative} to the total budget. The sum $(R\beta_e+D)/(\gramps+\ecal)$---the normalized, deterministic tendency $\partial_t(\gramps+\ecal)/(\gramps+\ecal)$---is shown as a dashed orange curve. Note that this tendency is positive at 10 and 20 km even though $\gramps+\ecal$ is actually decreasing. Without stochastic forcing, the system will always approach state $\va$ or $\vb$, depending on where the initial condition falls relative to the surface dividing the two attractors.

To quantify the critical role of stochastic forcing in effecting the transition at each committor level, we define the stochastic tendency of $\gramps+\ecal$ along transition paths:

\begin{align}
	\lcal_{AB}[\gramps+\ecal](\vx)&=\\
	&\hspace{-1.5cm}\lim_{\Delta t\to0}\ex\bigg[\frac{(\gramps+\ecal)(\bx(t+\Delta t))-(\gramps+\ecal)(\bx(t-\Delta t))}{2\Delta t} \nonumber\\
	&\hspace{-1.0cm}\Big|\,\bx(t)=\vx\text{ and $\bx(t)$ is in transition}\bigg] \nonumber\\
\end{align}
which is related to the ordinary infinitesimal generator $\lcal$ (see \citet{Oksendal} for mathematical background and the appendix of \citet{Finkel2021learning} for its application to the Holton-Mass model). The supplement describes the numerical procedure to approximate $\lcal_{AB}$ using short trajectories and a finite lag time. There, we show that $\lcal_{AB}f(\vx)$ is related to $\bj_{AB}\cdot\nabla f(\vx)$ for any observable $f$, so it is appropriate to view the arrows in Fig.~\ref{fig:projections_2d} and~\ref{fig:current_gebudget} as a proxy for the stochastic tendencies of the projected observables.

We introduce $\lcal_{AB}$ to compare with the deterministic tendency $\partial_t(\gramps+\ecal)(\vx)$, which for a diffusion process of the form~\eqref{eq:ito_diffusion} is simply $\bm v(\vx)\cdot\nabla(\gramps+\ecal)(\vx)$ by the chain rule. Their difference shows the impact of stochastic forcing responsible for transitions. More specifically, $\lcal_{AB}-\partial_t$ averaged over a committor level $q_0$ highlights the stochastic effects responsible for taking the system from $q_0$ to $q_0+dq$. Often it is not just a single coin flip that decides the fate of $\bx(t)$, but a whole sequence of random turns through state space aligning in just such a way to navigate from $A$ to $B$.  

The role of stochasticity is most stark at 10 and 20 km (panels (d) and (e)) and for $\qp_B<0.5$, where $\lcal_{AB}(\gramps+\ecal)$ is negative while $\partial_t(\gramps+\ecal)$ is positive, due to a strong positive tug of radiative cooling versus the weak dissipation of enstrophy. As $\qp_B$ increases, the stochastic and deterministic tendencies grow closer together: the more likely the transition to $B$, the easier it is for deterministic drift to carry it out alone. At 30 km (panel f), all forms of dissipation and forcing start out \emph{relatively} small compared to the magnitude of $\gramps+\ecal$, but as the path progresses they all diverge away from zero. Most notably, the stochastic and deterministic tendencies never diverge very far; if anything, stochastic noise \emph{slows} the collapse of $U$(30 km) at the end. It seems that to achieve the $A\to B$ transition, which is defined entirely in terms of $U$(30 km), the most common mechanism is a persistent negative push applied to lower altitudes, and this ultimately sets up the higher altitudes for more sudden, deterministic collapse after the ``hard work'' of eroding the vortex from below is mostly finished. 

In summary, the TPT diagnostics have demonstrated that the SSW process begins with steady, significant decay of the PV gradient (here, its squared gradient, $\gramps$) at lower altitudes, driven by the stochastic forcing, with only conservative changes taking place at higher altitudes. This preconditioning of the vortex opens up a valve to the mid-stratosphere. In the late stages of the transition, starting between $\qp_B=0.6$ and 0.7, the upper-level winds decline very suddenly. This begins conservatively as eddies grow, exchanging energy with the mean flow, and finishes non-conservatively, as friction dissipates the waves.

\section{Conclusion}
\label{sec:conclusion}

Transition path theory (TPT) is a mathematical framework that can be used to assess the near-term predictability and long-term climatology of anomalous weather events. The framework lends itself naturally to events associated with regime transitions, but it can be applied to more general anomalies. The key is to be able to define a suitable ``reaction coordinate'', or measure of progress, linking the event to the mean state. We have analyzed the statistical ensemble of Sudden Stratospheric Warmings (SSWs) in the idealized Holton-Mass model. Here, measures of the vortex strength (or the mean potential vorticity) and heat flux (eddy enstrophy) provide natural coordinates for applying the theory.

Probability densities and currents tell us how the system evolves through state space during a breakdown of the polar stratospheric vortex. The reactive current, $\bj_{AB}$, allows one to condition dynamical tendencies on the occurrence of a rare event. By overlaying $\bj_{AB}$ over observable subspaces at different altitudes in the stratosphere, we have identified the key roles of dissipation and stochastic forcing in driving SSWs in the Holton-Mass model. The stochastic driving represents the effects of unresolved Rossby and gravity waves that have been stripped from this highly truncated model. The action of these non-conservative processes, stochastic driving in particular, matter most at lower altitudes early in the transition process, conditioning the vortex, while the higher altitudes are shielded from significant dissipation. It is only late in the transition process, after the likelihood of the event has surpassed 60\%, that the upper-level winds play a significant role in the dynamics.

This work is an early application of TPT to atmospheric science. We believe it holds potential as a framework for forecasting, risk analysis, and uncertainty quantification. Thus far, it has been used mainly to analyze protein folding in molecular dynamics, but is now being applied in diverse fields such as social science \citep{Helfmann2021statistical}, as well as ocean and atmospheric science \citep{Finkel2020,Helfmann2020,Miron2020,Miron2022transition}. TPT results are best interpreted when viewed in a physically meaningful observable subspace of variables. Utilizing physical knowledge and experience with the system allows one to gain the most from the methodology. With the rather simple Holton-Mass model, we identified such a subspace based on an enstrophy budget. In different versions of quasigeostrophic dynamics, the wave activity \citep{Nakamura2010,Lubis2018role} and other diagnostics based on the transformed-Eulerian-mean \citep{Andrews1976planetary} are likely to be informative coordinates. 

Significant challenges remain for deploying TPT analysis at scale to state-of-the-art climate models. We have used a Dynamical Galerkin Approximation (DGA) short trajectory analysis algorithm to compute TPT quantities. One important limitation of this computational pipeline is the data generation step. We used a long direct simulation to sample the background climatology, which served the double purpose of seeding initial data points for short trajectories and providing a ground truth for validating the accuracy of DGA. The former point is critical: one must cover the space of initial conditions to capture the dynamics of extreme events. In some cases, short trajectory data already exist, e.g., from the subseasonal-to-seasonal (S2S) database \citep{Vitart2018s2s}, which we have used recently in \citet{Finkel2022revealing} to estimate centennial-scale SSW rates from only 21 years of ensemble forecasts. In other cases, it is advantageous to generate fresh data in undersampled regions of state space, which would require more advanced sampling methods such as the adaptive sampling strategies proposed in \cite{lucente2021coupling} and \cite{Strahan2022nnforecast}, or rare event simulation schemes such as in
\cite{Mohamad2018sequential}, \cite{Ragone2018heatwaves}, \cite{webber}, and \cite{Ragone2020averaged}.








\acknowledgments
During the time of writing, J.F. was supported by the U.S. DOE, Office
of Science, Office of Advanced Scientific Computing Research, Department of Energy Computational
Science Graduate Fellowship under Award Number DE-SC0019323. 
During the time of writing, R.J.W. was supported by 
New York University's Dean's Dissertation Fellowship and by
the Research Training Group in Modeling and Simulation funded by the NSF via grant RTG/DMS-1646339.
E.P.G. acknowledges support from the NSF through grants AGS-1852727 and OAC-2004572. This work was partially supported by the NASA Astrobiology Program, grant No.~80NSSC18K0829 and benefited from participation in the NASA
Nexus for Exoplanet Systems Science research coordination network. J.W. acknowledges support from the Advanced Scientific Computing Research Program within the DOE Office of Science through award DE-SC0020427 and from the NSF through award DMS-2054306. The computations in the paper were done on the high-performance computing cluster at New York University.

We thank John Strahan, Aaron Dinner, and Chatipat Lorpaiboon for many helpful conversations and methodological advice.

%
%
\datastatement
The code to produce the data set and results, either on the Holton-Mass model or on other systems, is publicly available at \url{https://github.com/justinfocus12/SHORT}. Interested users are encouraged to contact J.F. for more guidance on usage of the code.

%






%
%
%
\bibliographystyle{ametsoc2014_arxiv}
\bibliography{tpt_references.bib}

%

%

\end{document}


\maketitle

This document has three sections. Section \ref{sec:si_tpt_math} spells out transition path theory (TPT) formally, with definitions and equations for all quantities of interest without regard to their numerical approximation. Section \ref{sec:si_methods} describes the numerical method, dynamical Galerkin Approximation (DGA), and provides some numerical benchmarks. Section \ref{sec:si_minimum_action_method} gives details about the optimization method we used to find a minimum-action path. Finally, section \ref{sec:si_enstrophy_budget} gives a detailed derivation of the enstrophy budget presented in section 5 of the main manuscript.

\section{Transition path theory formalism} \label{sec:si_tpt_math}

We begin a quantitative description of transition paths by formalizing the notion of the transition path ensemble. The theoretical development parallels \cite{VandenEijnden2006}, but expands on it in several ways. Consider the stratosphere, or any other stochastic ergodic dynamical system, evolving through a very long time interval $(-T,T)$, during which it crosses from $A$ to $B$ and back a number $M_T$ of times. As $T\to\infty$, ergodicity guarantees that $M_T\to\infty$ as well. The $m$th transition path begins at time $\tau_m^-$ (so $\bx(\tau_m^-)\in A$\footnote{Technically, we assume $\bx(t)$ is right-continuous with left limits, meaning $\bx(\tau_m^-)\notin A$ but $\lim_{t\uparrow\tau_m^-}\bx(t)\in A$. This detail is not important for us here.}) and ends at time $\tau_m^+$ (so $\bx(\tau_m^+)\in B$). Each $\tau_m^-$ marks the beginning of an orange segment in Fig. 2, and $\tau_m^+$ marks the end of it. 

In principle, any statistical average over the transition path ensemble can be found by ``ergodic simulation'' (ES), in which we integrate the system for a long enough time to collect a large number of transition path samples. Although ES is simple and general, it is expensive for high-dimensional models, particularly for rare event simulation. The DGA method, explained below in section \ref{sec:si_methods}, circumvents ES by using only short trajectories (20 days long in our implementation). These are short not only compared to the return time $\tau_{m+1}^--\tau_m^+$ ($\sim1700$ days for the Holton-Mass model), but even compared to the $(A\to B)$ transit time $\tau_m^+-\tau_m^-$ ($\sim80$ days for the Holton-Mass model). DGA can be used to compute an important class of quantities including the committor functions, reactive density, and reactive current. The primitive ingredients of this calculation are \emph{forecast functions}, defined below. 

\subsection{Forecast functions}
The essential insight of TPT is to express the quantities of interest in terms of a set of \emph{forecast functions}. We consider a probabilistic forecast to be an estimate of the future conditioned on the present, of the general form 
\begin{align}
    F^+(\vx)=\ex_\vx[Q(\{(t,\bx(t)):t\geq0\})].
\end{align}
Here, $\ex_\vx$ indicates a conditional expectation given a fixed initial condition $\bx(0)=\vx$ (we can set $t_0=0$ when assuming autonomous dynamics). $Q$ is a generic functional of the future evolution of the state $\bx(t)$. It is explicitly a random variable under the stochastic forcing we impose here, but even in a deterministic model, uncertainty from initial conditions and model error lead to effective randomness. For example, $Q$ could return 1 if $\bx(t)$ next hits $B$ before $A$, and 0 if $\bx(t)$ next hits $A$ before $B$. This makes $F^+$ simply the forward committor, as introduced in section 3 of the main text:
\begin{align}
    F^+(\vx)&=\ex_\vx\big[\ind_B(\bx(\tau_{A\cup B}^+))\big]\\
    &=\pr_\vx\{\bx(\tau_{A\cup B}^+)\in B\}=:\qp_B(\vx)
\end{align}
We might also wish to forecast the time it takes to get there, by defining $Q=\tau_{A\cup B}^+\ind_B(\bx(\tau_{A\cup B}^+))$, which then gives us the expected lead time $\eta_B^+(\vx)=\ex_\vx[Q]/\qp_B(\vx)$. 

As explained in section (3) of the main text, the forward committor only looks to the future, and the backward committor is used to estimate where the system came from in the past: 
\begin{align}
    \qm_A(\vx)&=\ex_\vx[\ind_A(\bx(\tau_{A\cup B}^-))]=\pr_\vx\{\bx(\tau_{A\cup B}^-)\in A\}
\end{align}
This is a backward-in-time forecast, or \emph{aftcast}. 

Forward and backward committors are central components in the existing transition path theory laid out in \cite{E2006towards,VandenEijnden2006,tpt_mjp,tpt_simple_examples,pathfinding}, and elsewhere. Here, we generalize committors to forecast not only where the trajectory ends up, but what happens along the way. We consider forecast/aftcast functions of the form
\begin{align}
    \fp(\vx;\lambda)&=\ex_\vx\bigg[\ind_B(\bx(\tau_{A\cup B}^+))\exp\bigg(\lambda\int_0^{\tau_{A\cup B}^+}\dam(\bx(r))\,dr\bigg)\bigg] \label{eq:forecast}\\
    \fm(\vx;\lambda)&=\ex_\vx\bigg[\ind_A(\bx(\tau_{A\cup B}^-))\exp\bigg(\lambda\int_{\tau_{A\cup B}^-}^0\dam(\bx(r))\,dr\bigg)\bigg]
    \label{eq:aftcast}
\end{align}
where $\lambda$ is a real free parameter. For certain extreme weather events, $\dam$ might be chosen to measure accumulated damage of some kind, say, the total rainfall deposited over an area (in the case of hurricanes) or total time with surface temperatures above a certain threshold (in the case of heat waves). In a downward-coupled SSW model, one could define $\dam$ to reflect the human impact of extreme cold spells. However, in this paper we only ever set $\dam=1$, so the integral is the expected lead time.   

Everything we say about transition paths stems originally from the functions $\fp$ and $\fm$ for various $\dam$, as well as the steady-state distribution $\pi$. Thus, we will now express the quantities of interest in terms of $\pi$, $\fp$, $\fm$, and their $\lambda$-derivatives. Section \ref{sec:si_methods} will then explain how to compute them using short simulation data. 

%

\subsection{Ergodic averages}
The key to transforming forecasts into ensemble averages (at either level) is the \emph{ergodic assumption}, which goes as follows. Let $\by(t)$ denote all the hidden variables of the system responsible for apparent randomness, such as unresolved turbulence, so that the joint process $(\bx(t),\by(t))$ is completely deterministic. (In simulation, $\by(t)$ is the state of a pseudo-random number generator.) The hypothesis then states that there exists a probability density $p(\vx,\vy)$ such that for any bounded observable $G(\bx(t),\by(t))$, 
\begin{align}
	\label{eq:ergodicity_xy}
	\lim_{T\to\infty}\frac1T&\int_0^TG(\bx(t),\by(t))\,dt\\
	&=\int\bigg(\int G(\vx,\vy)p(\vx,\vy)\,d\vy\bigg)\pi(\vx)\,d\vx \nonumber\\
	&=:\int\dam(\vx)\pi(\vx)\,d\vx=:\ang{\dam}_\pi.
\end{align}
As a prototypical example, define
\begin{align}
	G(\vx,\vy)=\begin{cases} 1 & (\vx,\vy)\text{ comes from $A$ and goes to $B$ } \\ 0 & \text{otherwise} \end{cases}
\end{align}
where knowledge of $\vy$ lets us run the system forward and backward deterministically to evaluate the source and destination of $(\vx,\vy)$. In this case, Eq.~\eqref{eq:ergodicity_xy} becomes
\begin{align}
	\lim_{T\to\infty}\frac1T&\int_0^T\ind_A(\bx(\tau_{A\cup B}^-(t)))\ind_B(\bx(\tau_{A\cup B}^+(t))\,dt \nonumber\\
	&=\int\qm_A(\vx)\qp_B(\vx)\pi(\vx)\,d\vx=:\ang{\qm_A\qp_B}_\pi. \label{eq:timefrac_ab}
\end{align}
The left-hand side is the time fraction spent on the way from $A$ to $B$, and can be estimated from ES. On the right hand side, $\dam(\vx)=\qp_A(\vx)\qp_B(\vx)$. Substituting different combinations of $A$ and $B$ in Eq.~\eqref{eq:timefrac_ab} gives us the time fraction spend in the phases $B\to A$ ($\dam=\qm_B\qp_A$), $A\to A$ ($\dam=\qm_A\qp_A$), and $B\to B$ ($\dam=\qm_B\qp_B$). Fig.~\ref{fig:dga_validation}a shows these estimates from ES and DGA, which is further described in section \ref{sec:si_methods} below. 

\subsection{Transition path averages and currents}
In this section we use forecast functions~\eqref{eq:forecast} and~\eqref{eq:aftcast} to express the transition rate and reactive current. In fact, we will go slightly further and define a \emph{generalized rate}:
\begin{align}
    \rg(\lambda):&=\lim_{T\to\infty}\frac{1}{T}\sum_{m=1}^{M_T}\exp\bigg(\lambda\int_{\tau^-_m}^{\tau^+_m}\dam(\bx(r))\,dr\bigg) \label{eq:genrate_def}
\end{align}
The notation emphasizes that $\rg$ depends on the observable $\dam$ and the real parameter $\lambda$. To unpack this formula, first set $\lambda=0$ and observe that $\rg(0)=\frac{M_T}{T}$ is the number of transitions per unit time, or ordinary rate, whose inverse is the average period of the full SSW life cycle.\footnote{This is not to be confused with the asymmetric forward and backward rates,
\begin{align}
    k_{AB}=\frac{\rg(0)}{\ang{\qm_A}_\pi}, && 
    k_{BA}=\frac{\rg(0)}{\ang{\qm_B}_\pi}
\end{align}
which distinguish the $A\to B$ and $B\to A$ directions by how fast they occur. The factor $\ang{\qm_A}_\pi$ is the time fraction spent \emph{having last been in} $A$ rather than $B$, and $\ang{\qm_B}$ is the opposite. For example, if $A$ were very stable and $B$ very unstable, the system would spend most of its time in the basin of attraction of $A$, making $\ang{\qm_A}_\pi$ large and $k_{AB}\ll k_{BA}$. Asymmetric rates (or ``rate constants'') are very important for chemistry applications, but the symmetric rate is more useful to us presently.}

The ordinary rate has been studied extensively with TPT and preceding theories. A novel idea that we introduce here is to include the exponential factor $\exp\big(\lambda\int\dam(\bx(t))\,dt\big)$, though we do not present these results in this paper. The theoretical development below therefore reduces to classical TPT by replacing $\dam$ with 0. 

Returning to~\eqref{eq:genrate_def}, we divide through by $\rg(0)$:
\begin{align}
    \frac{\rg(\lambda)}{\rg(0)}&=\lim_{T\to\infty}\frac1{M_T}\sum_{m=1}^{M_T}\exp\bigg(\lambda\int_{\tau^-_m}^{\tau^+_m}\dam(\bx(r))\,dr\bigg) \nonumber\\
    &=\ex_{\mathrm{paths}}\bigg[\exp\bigg(\lambda\int_{\tau_A^-}^{\tau_B^+}\dam(\bx(r))\,dr\bigg)\bigg] \label{eq:transpathint_mgf}
\end{align}
where the subscript ``paths'' distinguishes the expectation as over all transition paths. The right side of~\eqref{eq:transpathint_mgf} is a moment-generating function for the integral $\int\dam(\bx(t))\,dt$ over the transition path. Differentiating in $\lambda$ yields the moments of the distribution of the integral, allowing one to calculate variance, skew, and kurtosis:
\begin{align}
    \frac{\partial_\lambda^k\rg(0)}{\rg(0)}&=\ex_\mathrm{paths}\bigg[\bigg(\int_{\tau_A^-}^{\tau_B^+}\dam(\bx(r))\,dr\bigg)^k\bigg], \label{eq:transpathint_moments}
\end{align}
Thus, $\rg(\lambda)$ contains much information about the transition ensemble.

We now express $\rg$ in terms of the forecast functions $\fp$ and $\fm$, again using the key assumption of ergodicity. We must convert Equation \eqref{eq:genrate_def}, a sum over transition paths $\sum_{m=1}^{M_T}(\cdot)$, into an integral over time $\int_{-T}^T(\cdot)\,dt$ and then (by ergodicity) into an integral over space $\int_{\R^d}(\cdot)\pi(\vx)\,d\vx$. This approach extends the rate derivation in \cite{VandenEijnden2006} and \cite{Strahan2020} to generalized rates. 

To write the rate as a time integral, we introduce a dividing surface between $A$ and $B$ (such as a committor level surface) and use the fact that a transition path crosses such a surface an odd number of times. A mask is applied to the time integral to select only the time segments when a reactive trajectory segment is crossing this surface ($+1$ for positive crossings and $-1$ for negative crossings), resulting in unit weight for each transition path. To be more explicit, let $S$ be a region of state space that contains $A$ and excludes $B$, so that its boundary $C=\partial S$ is a dividing surface between $A$ and $B$. The generalized rate~\eqref{eq:genrate_def} can then be written as the following time integral: 
\begin{align}
    \rg(\lambda)&=\lim_{\Delta t\to0}\frac1{\Delta t}\lim_{T\to\infty}\frac1T\int_{0}^T \label{eq:genrate0}\\
    &\exp\bigg(\lambda\int_{\tau_{A\cup B}^-(t)}^{\tau_{A\cup B}^+(t+\Delta t)}\dam(\bx(r))\,dr\bigg)\times\label{eq:genrate1}\\
    &\ind_A(\bx(\tau_{A\cup B}^-(t)))\ind_B(\bx(\tau_{A\cup B}^+(t+\Delta t)))\times \label{eq:genrate2}\\
    \Big[&\ind_S(\bx(t))\ind_{S\cm}(\bx(t+\Delta t))\label{eq:genrate3}\\
    -&\ind_{S\cm}(\bx(t))\ind_S(\bx(t+\Delta t))\Big]\,dt\label{eq:genrate4}
\end{align}
The idea is to restrict the interval $(0,T)$ to the collection of time intervals $(t,t+\Delta t)$ during which the path crosses the surface $\partial S$. Line \eqref{eq:genrate2} applies a mask picking out transition path segments, which are those that come from $A$ and next go to $B$. Line \eqref{eq:genrate3} applies a further mask picking out the narrow time intervals when $\bx(t)$ exits the region from $S$ to $S\cm$, while line \eqref{eq:genrate4} subtracts the backward crossings from $S\cm$ to $S$. Using ergodicity, we can replace the time integral with a space integral and insert conditional expectations inside. For example, the part of the integrand
\begin{align}
    \exp\bigg(\lambda&\int_{t+\Delta t}^{\tau_{A\cup B}^+(t+\Delta t)}\dam(\bx(r))\,dr\bigg)\times\\
    &\ind_B(\bx(\tau_{A\cup B}^+(t+\Delta t)))\ind_{S\cm}(\bx(t+\Delta t))\nonumber
\end{align}
becomes, after taking conditional expectations,
\begin{align}
    \ex\big[&\ind_{S\cm}(\bx(t+\Delta t))\fp(\bx(t+\Delta t))\big|\bx(t)=\vx\big]\\
    &=:\tcal^{\Delta t}\big[\ind_{S\cm}\fp\big](\vx)\nonumber
\end{align}
Where the \emph{transition operator} is defined as $\tcal^{\Delta t}f(\vx)=\ex_\vx[f(\bx(\Delta t))]$. Applying similar logic to all terms in the integrand, we have the following generalized rate formula: 
\begin{align}
    \rg&(\lambda)=\lim_{\Delta t\to0}\frac1{\Delta t}\int_{\R^d}\fm(\vx;\lambda)\times \label{eq:rate_formula_surface}\\
    &\Big\{\ind_S\tcal^{\Delta t}\big[\ind_{S\cm}\fp\big]
    -\ind_{S\cm}\tcal^{\Delta t}\big[\ind_S\fp\big]\Big\}(\vx)\pi(\vx)\,d\vx \nonumber
\end{align}
which holds for any $S$ enclosing $A$ and disjoint from $B$. This is a form estimable from short simulation data, which the next section will explain. 

The rate formula \eqref{eq:rate_formula_surface} is suggestive of a surface integral, counting hopping events across the surface $\partial S$. In fact, the reactive current $\bj_{AB}$ is defined as the vector field whose surface integral is equal to the symmetric rate:
\begin{align}
    \rg(0)=\int_{C}\bj_{AB}\cdot\vn\,d\sigma \label{eq:current_flux_relationship}
\end{align}
We have visualized $\bj_{AB}$ in sections 3 and 5 of the main text using a discretization of the integrand in~\eqref{eq:rate_formula_surface}. 

We have now completely described the mathematics of TPT, and our extensions to it. Exact knowledge of $\pi$, $\fp$, $\fm$, and their $\lambda$-derivatives is enough to generate all of the figures shown so far. The next section explains both how to compute these fundamental ingredients from data and assemble them into generalized rates. 

\section{Numerical method: dynamical Galerkin approximation (DGA)}
 \label{sec:si_methods}

\subsection{Feynman-Kac formulae}
We now sketch the numerical method, following \cite{dga,Strahan2020}, and \cite{Finkel2021learning}. Equations \eqref{eq:forecast} and \eqref{eq:aftcast} involve an integral in time all the way from $t=0$ to $t=\tau_{A\cup B}^+$ or (in backward time) to $\tau_{A\cup B}^-$, when $\bx(t)$ hits either $A$ or $B$ after wandering through state space for an indeterminate period. This would seem to require long trajectories to estimate. However, below we write $F^\pm_\dam$ as solutions to partial differential equations called Feynman-Kac formulae \cite{Oksendal,Karatzas,weinan2019applied}, which read
\begin{align}
    \begin{cases}
        (\lcal+\lambda\dam(\vx))\fp(\vx;\lambda)=0 & \vx\in D\\
        \fp(\vx)=0 & \vx\in A\\
        \fp(\vx)=1 & \vx\in B
    \end{cases} \label{eq:kbe_fwd}\\
    \text{ where }\lcal\phi(\vx):=\lim_{\Delta t\to0}\frac{\ex_\vx[\phi(\bx(\Delta t))]-\phi(\vx)}{\Delta t} \label{eq:gen_fwd}
    \\
    \begin{cases}
        (\wt\lcal+\lambda\dam(\vx))\fm(\vx;\lambda)=0 & \vx\in D\\
        \fm(\vx)=1 & \vx\in A\\
        \fm(\vx)=0 & \vx\in B
    \end{cases} \label{eq:kbe_bwd}\\
    \text{ where }\wt\lcal\phi(\vx):=\lim_{\Delta t\to0}\frac{\ex_\vx[\phi(\bx(-\Delta t))]-\phi(\vx)}{\Delta t} \label{eq:gen_bwd}
\end{align}
The linear operators $\lcal$ and $\wt\lcal$ are known as the forward and backward infinitesimal generators, pushing observable functions $\phi$ forward or backward in time analogously to a material derivative in fluid mechanics. The first term in the numerator of~\eqref{eq:gen_fwd} is the transition operator. The backward-in-time expectations are defined specifically for the \emph{equilibrium} process, leading to $\wt\lcal\phi(\vx)=\frac1{\pi(\vx)}\lcal\st[\pi\phi](\vx)$, where $\lcal\st$ is the adjoint of $\lcal$ with respect to the reference (Lebesgue) measure $d\vx$. Equivalently, $\wt\lcal$ is the adjoint of $\lcal$ with respect to the steady-state measure $\pi(\vx)\,d\vx$. In addition, we have the stationary Fokker-Planck equation for $\pi$ itself: 
\begin{align}
    \begin{cases}
        \lcal\st\pi(\vx)=0 & \vx\in\R^d\\
        \int_{\R^d}\pi(\vx)\,d\vx=1
    \end{cases}
    \label{eq:fpe}
\end{align}
We can further obtain equations for the derivatives of $F^\pm_\dam$ with respect to $\lambda$, using the Kac moment method \cite{Fitzsimmons1999kac} as also described in \cite{Finkel2021learning}. Differentiating Eq.~\eqref{eq:kbe_fwd} in $\lambda$ and setting $\lambda=0$ yields a recursive sequence of equations for the higher derivatives of $\fp$:
\begin{align}
    \lcal\big[\partial_\lambda^k\fp\big](\vx;0)&=-k\dam\partial_\lambda^{k-1}\fp(\vx;0), && k\geq1, \label{eq:kac_moment_method}
\end{align}
with boundary conditions $\partial_\lambda^k\fp\big|_{A\cup B}=0$. The same procedure can be applied to the aftcast $\fm$. 
Thus, our entire numerical pipeline boils down to solving equations of the
form~\eqref{eq:kbe_fwd},~\eqref{eq:kbe_bwd}, and~\eqref{eq:fpe}, as well as the inhomogeneous version~\eqref{eq:kac_moment_method}. In particular, the expected lead time $\tbd_B$ is equal to $\frac1{\qp_B(\vx)}\partial_\lambda F_\dam^+(\vx;0)$ where $\dam(\vx)\equiv1$. In other words,
\begin{align}
	\lcal\big[\qp_B\tbd_B\big](\vx)&=-\qp_B(\vx)
\end{align}
and the expected lead time $\tbd_B$ can only be computed after first computing the committor $\qp_B$.

 
Section 5 additionally defined the \emph{stochastic tendency} $\lcal_{AB}\phi$ for any observable $\phi$. The idea is to turn the expectations in Eqs.~\eqref{eq:gen_fwd} and~\eqref{eq:gen_bwd} into conditional expectations, given that a transition is underway through $\vx$. As such, $\lcal_{AB}$ is defined so that 

\begin{align}
	\lcal_{AB}\phi(\vx)&=\lim_{\Delta t\to0}\ex_\vx\bigg[\frac{\phi(\bx(\Delta t))-\phi(\bx(-\Delta t))}{2\Delta t}\bigg|\bx(\tau_{A\cup B}^-)\in A\text{ and }\bx(\tau_{A\cup B}^+)\in B\bigg] 
\end{align}
To reveal its connection with the reactive current, and to justify the discretization to follow, we manipulate this formula to express $\lcal_{AB}$ in terms of committors. Above, $\tau_{A\cup B}^+$ is shorthand for the first hitting time to $A\cup B$ after $t=0$, and $\tau_{A\cup B}^-$ is shorthand for the most recent hitting time to $A\cup B$ before $t=0$. In the following, we abbreviate $\tau_{A\cup B}^\pm$ by $\tau^\pm$ to reduce clutter, and use the fact that $\tau^+(0)=\tau^+(\Delta t)$ and $\tau^-(0)=\tau^-(-\Delta t)$ for small enough $\Delta t$ and for $\vx$ sufficiently far from $A\cup B$. 
\begin{align}
	\lcal_{AB}\phi(\vx)=&\lim_{\Delta t\to0}\frac1{2\Delta t}\frac{\ex_\vx\big[\ind_A(\bx(\tau^-))\ind_B(\bx(\tau^+))\big(\phi(\bx(\Delta t))-\phi(\bx(-\Delta t))\big)\big]}{\ex_\vx\big[\ind_A(\bx(\tau^-))\ind_B(\bx(\tau^+))\big]}\\
	=\frac1{\qm_A(\vx)\qp_B(\vx)}
	&\lim_{\Delta t\to0}\ex_\vx\bigg[\ind_A(\bx(\tau^-))\ind_B(\bx(\tau^+))\frac{\phi(\bx(\Delta t))-\phi(\bx(-\Delta t))}{2\Delta t}\bigg]\\
	\label{eq:LAB_discretization}
	=\frac1{\qm_A(\vx)\qp_B(\vx)}
	\frac12\bigg\{
	&\lim_{\Delta t\to0}\ex_\vx\bigg[\ind_A\Big(\bx\big(\tau^-(0)\big)\Big)\ind_B\Big(\bx\big(\tau^+(\Delta t)\big)\Big)\frac{\phi(\bx(\Delta t))-\phi(\vx)}{\Delta t}\bigg] \\
	+&\lim_{\Delta t\to0}\ex_\vx\bigg[\ind_A\Big(\bx\big(\tau^-(-\Delta t)\big)\Big)\ind_B\Big(\bx\big(\tau^+(0)\big)\Big)\frac{\phi(\vx)-\phi(\bx(-\Delta t))}{\Delta t}\bigg]\bigg\} \nonumber	\\
	=\frac1{\qm_A(\vx)\qp_B(\vx)}
	\frac12\bigg\{
	&\lim_{\Delta t\to0}\ex_\vx\bigg[\ind_A\Big(\bx\big(\tau^-(0)\big)\Big)\frac{\ind_B(\bx(\tau^+(\Delta t)))\phi(\bx(\Delta t))-\ind_B(\bx(\tau^+(0)))\phi(\vx)}{\Delta t}\bigg]\\
	+&\lim_{\Delta t\to0}\ex_\vx\bigg[\ind_B\Big(\bx\big(\tau^+(0)\big)\Big)\frac{\ind_A(\bx(\tau^-(0)))\phi(\vx)-\ind_A(\bx(\tau^-(-\Delta t)))\phi(\bx(-\Delta t))}{\Delta t}\bigg] \nonumber
	\bigg\} \\
	=\frac1{\qm_A(\vx)\qp_B(\vx)}\frac12&\bigg\{\qm_A(\vx)\lcal\big[\qp_B\phi\big](\vx)-\qp_B\wt\lcal\big[\qm_A\phi\big](\vx)\bigg\} \\
	=&\frac12\bigg\{\frac{\lcal[\qp_B\phi](\vx)}{\qp_B(\vx)}-\frac{\wt\lcal[\qm_A\phi](\vx)}{\qm_A(\vx)}\bigg\} 
\end{align}

Now consider the average of $\lcal_{AB}\phi(\vx)$ with respect to the reactive density $\pi_{AB}$ restricted to a region $\mathscr{R}$ of state space, e.g., near a certain level set of the committor. We can express such an average as

\begin{align}
	\ang{\lcal_{AB}\phi}_{\pi_{AB}|\mathscr{R}}&=\frac{\int_{\mathscr{R}}\lcal_{AB}\phi(\vx)\qm_A(\vx)\qp_B(\vx)\pi(\vx)\,d\vx}{\int_{\mathscr{R}}\qm_A(\vx)\qp_B(\vx)\pi(\vx)\,d\vx}
	=\frac12\frac{\int_{\mathscr{R}}\big(\qm_A\lcal[\qp_B\phi]-\qp_B\lcal[\qm_A\phi]\big)\pi\,d\vx}{\int_{\mathscr{R}}\qm_A\qp_B\pi\,d\vx}
\end{align}
In \cite{Strahan2020}, Eq. S12, it is shown that 
\begin{align}
	\bj_{AB}\cdot\nabla\phi&=\frac12\Big(\qm_A\lcal\big[\qp_B\phi\big]-\qp_B\wt\lcal\big[\qm_A\phi\big]\Big)\pi
\end{align}
which means that 
\begin{align}
	\ang{\lcal_{AB}\phi}_{\pi_{AB}|\mathscr{R}}&=\frac{\int_{\mathscr{R}}\bj_{AB}\cdot\nabla\phi\,d\vx}{\int_{\mathscr{R}}\qm_A\qp_B\pi\,d\vx}
\end{align}
In words, this ratio conveys the total flow of reactive current across contours of $\phi$ in a region $\mathscr{R}$, adjusted for the likelihood for transition paths to be in $\mathscr{R}$ to begin with. The denominator ensures that regions with slow and fast movement of transition paths are compared fairly.

\subsection{Discretization of Feynman-Kac formulae}
We will now describe how to discretize and solve these three equations, which requires three similar but distinct procedures. 

First we attack~\eqref{eq:kbe_fwd}. The generator of a diffusion processes can be expressed as a partial differential operator, and so the above equations are PDEs over state space. PDEs cannot be practically discretized in high dimensions, but the essential property of spatial locality allows for data-driven approximation with short trajectories, using the probabilistic definition in \eqref{eq:gen_fwd} and \eqref{eq:gen_bwd}. This is how we use our large data set of short trajectories,
\begin{align}
    \{\bx_n(t):0\leq t\leq\Delta t\}_{n=1}^N,
\end{align}
where the initial points $\bx_n(0)$ are drawn from a \emph{sampling measure} $\mu$, which we will define in the following subsection. To discretize Eq.~\eqref{eq:kbe_fwd}, we first eliminate the numerically problematic limit $\Delta t\to0$ and integrate the equation using Dynkin's Formula \cite{Oksendal,weinan2019applied}: for any stopping time $\theta>0$, 
\begin{align}
    \ex_\vx[f(\bx(\theta))]&=f(\vx)+\ex_\vx\bigg[\int_0^\theta\lcal f(\bx(t))\,dt\bigg] \label{eq:dynkin}
\end{align}
Here we take $\theta=\min(\Delta t,\tau_{A\cup B})$. In other words, we artificially halt the $n$th trajectory $\bx_n(t)$ if it wanders into $A\cup B$ before the terminal time $\Delta t$. The $n$th stopping time from the data set is called $\theta_n$. The operator on the left-hand side of~\eqref{eq:dynkin} is known as the \emph{stopped} transition operator $\tcal^\theta$. Applying it to the unknown forecast function $\fp$ in~\eqref{eq:kbe_fwd} and using the fact $\lcal\fp=-\lambda\dam\fp$, we get
\begin{align}
    \tcal^\theta\fp(\vx;\lambda)&=\fp(\vx;\lambda)-\lambda\ex_\vx\bigg[\int_0^\theta\dam(\bx(t))\fp(\bx(t);\lambda)\,dt\bigg]
\end{align}
To be more concise, we define the integral operator $\kcal^\theta_\dam f(\vx)=\ex_\vx\big[\int_0^\theta\dam(\bx(t))f(\bx(t))\,dt\big]$, and write an integrated version of~\eqref{eq:kbe_fwd}:
\begin{align}
    \begin{cases}
        (\tcal^\theta-1+\lambda\kcal^\theta_\dam)\fp(\vx;\lambda)=0 & \vx\in D\\
        \fp(\vx;\lambda)=0 & \vx\in A \\
        \fp(\vx;\lambda)=1 & \vx\in B
    \end{cases}
    \label{eq:kbe_bwd_dynkin}
\end{align}
To discretize this equation and impose regularity on the solution, we approximate $\fp$ as a finite linear combination with coefficients $w_j(\fp(\vx;\lambda))$, which we abbreviate $w_j(\lambda)$ for simplicity:
\begin{align}
    \fp(\vx;\lambda)&\approx\hat{F}_\dam^+(\vx;\lambda)+\sum_{j=1}^Mw_j(\lambda)\phi_j(\vx;\lambda)
\end{align}
where $\hat{F}_\dam^+$ is a guess function obeying the boundary conditions on $A\cup B$, and $\{\phi_j\}_{j=1}^M$ is a collection of basis functions that are zero on $A\cup B$, which will be defined in the following subsection. The task is now to solve for the coefficients $w_j(\lambda)$. Equation~\eqref{eq:kbe_bwd_dynkin} becomes a system of linear equations in $w_j(\lambda)$:
\begin{align}
    \sum_{j=1}^Mw_j(\lambda)(\tcal^\theta-1+\lambda\kcal_\dam^\theta)\phi_j(\vx;\lambda)=-(\tcal^\theta-1+\lambda\kcal_\dam^\theta)\hat{F}_\dam^+(\vx;\lambda)
    \label{eq:collocation}
\end{align}
Since the transfer and integral operators are expectations over the future state of the system beginning at $\vx$, we can estimate their action at $\vx=\bx_n(0)$ (a short-trajectory starting point) as
\begin{align}
    (\tcal^\theta-1+\lambda\kcal_\dam^\theta)\phi_j(\bx_n(0);\lambda)&\approx\phi_j(\bx_n(\theta_n);\lambda)-\phi_j(\bx_n(0);\lambda) \label{eq:dga_core0} \\
    &\hspace{1cm}+\lambda\int_0^{\theta_n}\dam(\bx_n(t))\phi_j(\bx_n(t);\lambda)\,dt \nonumber
\end{align}
or, if multiple independent trajectories are launched from $\vx$, we can average over them. Applying this to every short trajectory and plugging into Eq.~\eqref{eq:collocation}, we obtain a system of $N$ equations in $M$ unknowns. In practice, $N\gg M$, meaning we have many more trajectories than basis functions, and the system is overdetermined. A unique, and regularized, solution is obtained by casting it into weak form: we multiply both sides by $\phi_i(\vx)$ and integrate over state space:
\begin{align}
    \sum_{j=1}^Mw_j(\lambda)\Big\langle\phi_i,(\tcal^\theta-1+\lambda\kcal_\dam^\theta)\phi_j\Big\rangle_\zeta&=-\Big\langle\phi_i,(\tcal^\theta-1+\lambda\kcal_\dam^\theta)\hat{F}_\dam^+\Big\rangle_\zeta
    \label{eq:dga_core1}
\end{align}
where the inner products are defined with respect to a measure $\zeta$:
\begin{align}
    \ang{f,g}_\zeta&=\int f(\vx)g(\vx)\zeta(\vx)\,d\vx
    \label{eq:dga_core2}
\end{align}
With our finite data set, we approximate the inner product by a sum over pairs of points. Given that $\bx_n(0)\sim\mu$, the law of large numbers ensures that for any bounded function $H(\vx)$, 
\begin{align}
    \frac1N\sum_{n=1}^NH(\bx_n(0))\approx\int H(\vx)\mu(\vx)\,d\vx
    \label{eq:dga_core3}
\end{align}
becomes more accurate as $N\to\infty$. Thus we set $H(\vx)=\phi_i(\vx)(\tcal^\theta-1+\lambda\kcal_\dam^\theta)\phi_j(\vx)$ as estimated by~\eqref{eq:dga_core0}, approximate the inner products with $\zeta=\mu$, plug them into~\eqref{eq:dga_core1}, and solve the $M\times M$ system of linear equations for $w_j(\lambda)$. 

Next we address~\eqref{eq:fpe}. The integrated version of~\eqref{eq:fpe} is found by observing that for any bounded function $H$,
\begin{align}
    \ex_{\bx(0)\sim\pi}[H(\bx(\Delta t))]=\ex_{\bx(\Delta t)\sim\pi}[H(\bx(\Delta t))]
    \label{eq:stationarity}
\end{align}
where $\bx(0)\sim\pi$ means the initial condition is drawn from equilibrium, and thus so is $\bx(\Delta t)$ since $\pi$ is stationary. Of course, our initial data is \emph{not} distributed according to $\pi$, but rather by $\mu$; the goal is to solve for the \emph{change of measure} $\chom(\vx)$. Writing~\eqref{eq:stationarity} as an integral,
\begin{align}
    \int\tcal^{\Delta t}H(\vx)\pi(\vx)\,d\vx&=\int H(\vx)\pi(\vx)\,d\vx\\
    0&=\int(\tcal^{\Delta t}-1)H(\vx)\pi(\vx)\,d\vx\\
    &=\int(\tcal^{\Delta t}-1)H(\vx)\chom(\vx)\mu(\vx)\,d\vx\\
    &=\bigg\langle(\tcal^{\Delta t}-1)H,\chom\bigg\rangle_\mu
\end{align}
As this holds for every bounded $H$, we enforce the equation for $H=\phi_i$, $i=1,\hdots,M$ and approximate $\chom=\sum_{j=1}^Mw_j\big(\chom\big)\phi_j$, resulting in a homogeneous linear system for the coefficients $w_j$ similar to~\eqref{eq:dga_core1}. The matrix elements are
\begin{align}
    \ang{(\tcal^{\Delta t}-1)\phi_i,\phi_j}_\mu=\ang{\phi_i,(\tcal^{\Delta t}-1)_\mu\st\phi_j}_\mu
\end{align}
where $(\cdot)_\mu\st$ denotes the adjoint operator with respect to $\mu$. We solve this homogeneous system by $QR$ decomposition. Note that there are no boundary conditions, and the trajectories need not be stopped early. Instead there is a normalization condition, which we enforce as $\sum_{n=1}^N\chom(\bx_n(0))=1$. If we were to divide by $\Delta t$ and take the limit $\Delta t\to0$, we would recover the strong form of the Fokker-Planck equation,~\eqref{eq:fpe}.

We use a particularly simple basis consisting of indicator functions,
\begin{align}
	\phi_j(\vx)=\ind_{S_j}(\vx)=\begin{cases}1 & \vx\in S_j \\ 0 & \vx\notin S_j\end{cases}
\end{align}
where $\{S_1,\hdots,S_M\}$ is a disjoint partition of state space. We obtain these regions by clustering data points with a hierarchical version of the K-means algorithm implemented in \texttt{scikit-learn} \cite{scikit-learn}, with $M=1500$. With such a basis, the inner products yield the entries of the Markov matrix $P_{ij}$ described in the main text. This guarantees a null space automatically. In a more general basis, one can add a constant vector to ensure a nontrivial null vector. 

Given the weights $\chom(\bx_n(0))$, we can take any ergodic average $\dam$ by inserting the change of measure:
\begin{align}
	\label{eq:numerical_ergodic_integral}
    \ang{\dam}_\pi&=\int_{\R^d}\dam(\vx)\pi(\vx)\,d\vx=\int_{\R^d}\dam(\vx)\,d\vx\approx\sum_{n=1}^N\dam(\bx_n(0))\chom(\bx_n(0))
\end{align}

For the specific case of a Markov state model, we can estimate the $\pi$-weighted state-space integral  by decomposing it over clusters:
\begin{align}
	\ang{\dam}_\pi&=\sum_{j=1}^M\int_{S_j}\dam(\vx)\pi(\vx)\,d\vx\\
	&\approx\sum_{j=1}^Mw_j(\pi)\times(\text{average of $\dam$ over $S_j$})\\
	&\approx\sum_{j=1}^Mw_j(\pi)\frac{\sum_{n=1}^N\dam(\bx_n(0))\ind_{S_j}(\bx_n(0))}{\sum_{n'=1}^N\ind_{S_j}(\bx_{n'}(0))}\\
	&=\sum_{n=1}^N\dam(\bx_n(0))\frac{w_{j(n)}(\pi)}{\#\{n':j(n')=j(n)\}}
\end{align}
where $j(n)$ is defined as the cluster that $\bx_{n}(0)$ is assigned to, i.e., $\bx_{n}(0)\in S_{j(n)}$. Hence the change of measure for a Markov state model is the coefficient of $\dam(\bx_n(0))$ in  the last equation. Note that it sums to one over all data points, as it must.

Finally, we address the time-reversed Kolmogorov equation~\eqref{eq:kbe_bwd}. The only modification from~\eqref{eq:kbe_fwd} is that the inner products in~\eqref{eq:dga_core1} are interpreted in backward time, i.e., with all trajectories reversed, $\bx_n(\Delta t)$ becoming the beginning and $\bx_n(0)$ becoming the end. The problem is that $\bx_n(\Delta t)$ is not distributed according to $\mu$, and so we cannot use the same Monte Carlo inner product as in Eq.~\eqref{eq:dga_core3} with reference measure $\zeta=\pi$. However, we can solve the problem by reweighting with the change of measure as follows, leading to $\zeta=\pi$. We let the trajectory be discrete in time, i.e.,
\begin{align}
    \bx_n=\bigg[\bx_n(0),\bx_n\bigg(\frac{\Delta t}{K}\bigg),\bx_n\bigg(\frac{2\Delta t}{K}\bigg),\hdots,\bx_n(\Delta t)\bigg]
\end{align}
and consider functionals $H[\bx_n]$ of the whole trajectory. Defining the transition density $p(\vx,\vy)$ for each step of size $\Delta t$, the expectation of $H$ with $\bx_n(0)\sim\pi$ is given by
\begin{align}
    \ex_{\bx(0)\sim\pi}H[\bx]=\int\,d\vx_0\pi(\vx_0)\int d\vx_1p(\vx_0,\vx_1)\int\hdots\int d\vx_Kp(\vx_{K-1},\vx_K)H[\vx_0,\hdots,\vx_K]
\end{align}
The time reversal step explicitly assumes the \emph{equilibrium} backward process, leading to a backward transition kernel $\wt p(\vy,\vx)=\frac{\pi(\vx)}{\pi(\vy)}p(\vx,\vy)$. Inserting this throughout converts the expectation over $\bx(0)$ into an expectation over $\bx(\Delta t)$:
\begin{align}
    \ex_{\bx(0)\sim\pi}H[\bx]&=\int d\vx_0\pi(\vx_0)\int d\vx_1\frac{\pi(\vx_1)}{\pi(\vx_0)}\wt p(\vx_1,\vx_0)\int\\
    &\hdots\int d\vx_K\frac{\pi(\vx_K)}{\pi(\vx_{K-1})}\wt p(\vx_K,\vx_{K-1})H[\vx_0,\hdots,\vx_K]\\
    &=\int d\vx_K\pi(\vx_K)\int d\vx_{K-1}\wt p(\vx_K,\vx_{K-1})\int\hdots\int d\vx_0\wt p(\vx_1,\vx_0) H[\vx_0,\hdots,\vx_K]\\
    &=\wt\ex_{\bx(\Delta t)\sim\pi}H[\bx]
\end{align}
where $\wt\ex$ denotes backward-in-time expectation. This is precisely what we need to apply~\eqref{eq:dga_core3} to the time-reversed process, namely, define $H$ such that
\begin{align}
    \phi_i(\bx(\Delta t))(\tcal^\theta-1+\lambda\kcal_\dam^\theta)\phi_j(\bx(\Delta t))&=:H[\bx]
\end{align}
and then integrate over state space weighted by $\pi$, turning the left-hand side into an inner product:
\begin{align}
    \ang{\phi_i,(\wt\tcal^{\theta}-1+\lambda\wt\kcal_\dam^\theta)\phi_j}_\pi&=\wt\ex_{\bx(\Delta t)\sim\pi}H[\bx]\\
    &=\ex_{\bx(0)\sim\pi}H[\bx]\approx\sum_{n=1}^NH[\bx_n]\chom(\bx_n(0))
\end{align}
The right-hand side of Eq.~\eqref{eq:dga_core3} can be estimated similarly, also with $\zeta=\pi$. 

In both forward- and backward-time estimates, we never solve for $\fp(\vx;\lambda)$ or $\fm(\vx;\lambda)$ with nonzero $\lambda$; rather, we repeat the recursion process with Eq.~\eqref{eq:kac_moment_method}. This is equivalent to implicitly differentiating the discretized system Eq.~\eqref{eq:dga_core1}. 

\subsection{Rate estimate and numerical benchmarking}
To estimate generalized rates (in particular, the ordinary rate), we reproduce here the rate estimate from \cite{Strahan2020} for reference, which is an almost-direct implementation of the formula~\eqref{eq:rate_formula_surface}, repeated here:
\begin{align}
    \rg&(\lambda)=\lim_{\Delta t\to0}\frac1{\Delta t}\int_{\R^d}\fm(\vx;\lambda)\times \label{eq:rate_formula_surface_repeat}\\
    &\Big\{\ind_S\tcal^{\Delta t}\big[\ind_{S\cm}\fp\big]
    -\ind_{S\cm}\tcal^{\Delta t}\big[\ind_S\fp\big]\Big\}(\vx)\pi(\vx)\,d\vx \nonumber
\end{align}
In principle, the integral could be estimated directly with any choice of dividing surface $S$, but the sum would only use data either exiting $S$ (first term) or entering $S$ (second term). We can use all the data at once and improve numerical stability by averaging over multiple such surfaces, and furthermore converting the transition operator $\tcal^{\Delta t}$ into the generator $\lcal$. However, we cannot simply take the limit under the integral due to the discontinuity in $\ind_S$. Instead we get a smooth function into the integrand with the following steps. First, replace $\ind_S$ with $1-\ind_{S\cm}$ everywhere:
\begin{align}
    \rg(\lambda)&=\lim_{\Delta t\to0}\frac1{\Delta t}\int_{\R^d}\fm(\vx;\lambda)\times\\
    &\Big\{(1-\ind_{S\cm})\tcal^{\Delta t}\big[\ind_{S\cm}\fp\big]-\ind_{S\cm}\tcal^{\Delta t}\big[(1-\ind_{S\cm})\fp\big]\Big\}(\vx)\pi(\vx)\,d\vx\\
    &=\lim_{\Delta t\to0}\frac1{\Delta t}\int_{\R^d}\fm(\vx;\lambda)\Big\{\tcal^{\Delta t}\big[\ind_{S\cm}\fp\big]-\ind_{S\cm}\tcal^{\Delta t}\fp\Big\}(\vx)\pi(\vx)\,d\vx\\
\end{align}
Next, add and subtract $\ind_{S\cm}\fp$ inside the integrand.
\begin{align}    
    \rg(\lambda)&=\lim_{\Delta t\to0}\int_{\R^d}\fm(\vx;\lambda)\bigg\{\frac{\tcal^{\Delta t}-1}{\Delta t}\big[\ind_{S\cm}\fp\big]-\ind_{S\cm}\frac{\tcal^{\Delta t}-1}{\Delta t}\fp\bigg\}(\vx)\pi(\vx)\,d\vx
\end{align}
At this point it is tempting to take the limit inside the integral, as $(\tcal^{\Delta t}-1)/\Delta t$ formally approaches $\lcal$. But the first term acts on a discontinuous function, which won't have a well-defined time derivative. We first replace $\ind_{S\cm}$ with a smooth function (on $D$), as follows. 

Let $K:\R^d\to[0,1]$ be a function that increases from 0 on set $A$ to 1 on set $B$ (for instance, the committor). Let $S_\zeta=\{\vx:K(\vx)\leq\zeta\}$ for $\zeta\in(0,1)$, and integrate both sides over $\zeta$, noting that $\int_0^1\ind_{S_\zeta\cm}(\vx)\,d\zeta=\int_0^1\ind\{K(\vx)>\zeta\}\,d\zeta=K(\vx)$. 
\begin{align}
    \int_0^1\rg(\lambda)\,d\zeta&=\lim_{\Delta t\to0}\int_{\R^d}\fm(\vx;\lambda)\bigg\{\frac{\tcal^{\Delta t}-1}{\Delta t}\big[K\fp\big]-K\lcal\fp\bigg\}(\vx)\pi(\vx)\,d\vx
\end{align}
Now we can move the limit inside and use the PDE to find
\begin{align}
    \rg(\lambda)&=\int_{\R^d}\fm(\vx;\lambda)\big\{\lcal[K\fp](\vx)+\lambda K(\vx)\dam(\vx)\fp(\vx)\big]\big\}\pi(\vx)\,d\vx
\end{align}
This formula can be estimated directly from knowledge of $\fm$,$\fp$, and $\pi$, using the ergodic assumption and with a discrete finite difference in time to estimate $\lcal[K\fp]$, i.e.,
\begin{align}
    \lcal[K\fp](\bx_n(0))\approx\frac{K(\bx_n(\Delta t))\fp(\bx_n(\Delta t))-K(\bx_n(0))\fp(\bx_n(0))}{\Delta t}
\end{align}
Derivatives with respect to $\lambda$ can be found by iterating the product rule, as we have solved for the derivatives of $\fp$ and $\fm$. 

To validate DGA numerically, we can compare to the results of ES. In \cite{Finkel2021learning} (Fig. 7), we saw convergence of the DGA committor to the ES committor across state space as sample size and lag time were increased. Here, we turn our attention to summary statistics of interest for full transition paths, not just forecasting. This will benchmark our current DGA implementation for comparison with future algorithmic developments.

\begin{figure}
	    \includegraphics[trim={5cm 0 15cm 0}, clip,width=0.45\linewidth]{"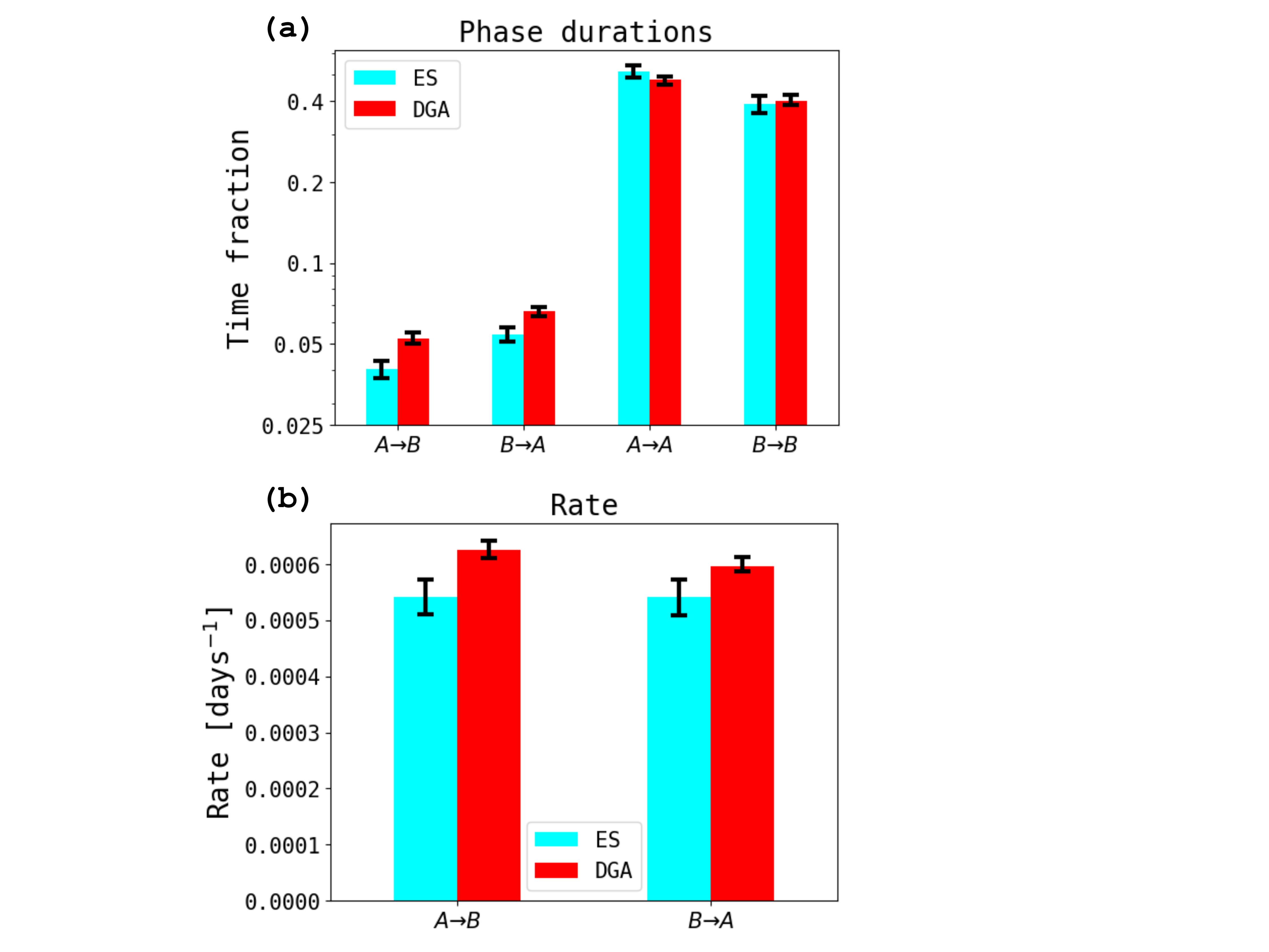"}
    \caption{\textbf{DGA benchmarks and comparison to ES.} (a) Time fractions spent in each phase, as calculated by DGA (red) and ES (cyan). (b) Total SSW rate estimated using both $\bj_{AB}$ and $\bj_{BA}$; the two cyan columns are identical, as the counting method includes one $B\to A$ transition for every $A\to B$ transition.} 
    \label{fig:dga_validation}
\end{figure}

Fig.~\ref{fig:dga_validation}a displays the time fractions spent in each phase of the SSW lifecycle: $A\to B$, $B\to A$, $A\to A$, and $B\to B$, including estimates from ES (cyan) and DGA (red) and their uncertainties. The DGA estimate of the $A\to B$ time fraction is a $\pi$-weighted average of $\qm_A(\vx)\qp_B(\vx)$ over state space,
\begin{align}
	\ang{\qm_A\qp_B}_\pi=\int\qm_A(\vx)\qp_B(\vx)\pi(\vx)\,d\vx
\end{align}
and similarly for the other phases (section \ref{sec:si_tpt_math} above justifies this formula rigorously, and section \ref{sec:si_methods} above details the numerical computation of the integral). The DGA error bars are generated by repeating the entire pipeline three times with different short trajectory realizations. The bar height shows the mean, and the error bars show the minimum and maximum. The ES error bars are generated by bootstrap resampling (with replacement) 500 times from the control simulation, treating an entire SSW lifecycle as a single unit (from the beginning of one $A\to B$ transition until the beginning of the next one). This assumes no memory between successive events, which we have found to be reasonable; there is insignificant autocorrelation between consecutive return periods. The bars extend two root-mean-squared errors in both directions, enclosing a 95\% confidence interval. To first order, DGA agrees well with ES on the fraction of time spent in each phase. $A$ is the more stable of the two regimes, accounting for $\sim50\%$ of the time compared to the $\sim40\%$ of time spent in the orbit of $B$. The transition events are both an order of magnitude shorter, with $B\to A$ taking slightly longer on average. DGA ranks the $A\to B$ and $B\to A$ time fractions correctly, despite a bias in the absolute magnitudes.

The numbers in Fig.~\ref{fig:dga_validation}a are only relative durations; they do not tell us how long a full life cycle takes. That number is given by (one over) the rate. Fig.~\ref{fig:dga_validation}b shows three different rate estimates (that is, the generalized rate with $\dam=0$) using the formulas above. The cyan bars come from ES, counting the number of $A\to B$ transitions per unit time. Of course, this equals the number of $B\to A$ transitions per unit time, so the $A\to B$ and $B\to A$ cyan bars are identical. Error bars come from bootstrapping, as with the relative durations. The red bars come from DGA, and these estimates are not technically symmetric. The DGA estimate labeled $A\to B$ integrates $\bj_{AB}\cdot\vn$ over dividing surfaces with $\vn$ pointing away from $A$ toward $B$, while the estimate labeled $B\to A$ integrates $\bj_{BA}\cdot\vn$ over surfaces with $\vn$ pointing away from $B$ toward $A$. Numerical and sampling errors cause slight differences between them, but Fig.~\ref{fig:dga_validation}b shows them both to come within 20\% of the ES estimate. 

DGA estimates should converge with increasing $M$ (cluster number) and $N$ (short-trajectory ensemble size). Larger $M$ makes the approximation space $\{\phi_1,\hdots,\phi_M\}$ more expressive, making finer estimates possible. However, as $M$ grows, we need more short trajectories $N$ to robustly estimate the entries of the expanding matrix ~\eqref{eq:dga_core1}. Conversely, as $M$ shrinks, $P_{ij}$ will become closer to diagonal, because trajectories will escape from their starting cluster less frequently. Thus $\Delta t$ would have to increase when $M$ decreases. The optimal choice for a given model will depend on the relative costs of integrating the model, building basis sets, and solving large linear systems on different computer architectures. With our choice of $M=1500$, increasing $N$ from $5\times10^4$ to $3\times10^6$ does not change the DGA point estimates very much, but shrinks the error bars by a factor of $\sim$4. To further reduce the bias in Fig.~\ref{fig:dga_validation}, we would likely need more refined basis functions, perhaps using nonlinear features as input to K-means. We do not yet have theoretical guarantees or optimal prescriptions for DGA parameters, but given the flexibility and parallelizability of the method, we believe it has much room for growth.

\subsection{Visualization method}
The two-dimensional projections in the main paper are generated with the following procedure. Let $\vy=\by(\vx)$ be an observable subspace, typically with dimension much less than that of $\vx$ (usually two). Any scalar field $F(\vx)$, such as the committor, has a projection $F^\by(\vy)$ onto this subspace by
\begin{align}
    F^\by(\vy)=\int F(\vx)\pi(\vx)\delta(\by(\vx)-\vy)\,d\vx
\end{align}
In practice, the $\vy$ space is partitioned into grid boxes $d\vy$, and the integral is estimated from the dataset, yielding
\begin{align}
	\label{eq:discrete_ergodic_average}
    F^\by(\vy)=\frac1N\sum_{n=1}^NF(\bx_n(0))\der\pi\mu(\bx_n(0))\ind_{d\vy}(\by(\bx_n(0)))
\end{align}
where $\ind_{d\vy}(\by(\vx))=1$ if $\by(\vx)\in\,d\vy$ and zero otherwise. In words, we simply take a weighted average over all data points $\bx_n(0)$ that project onto the grid box $d\vy$, with weights given by the change of measure. In Fig. 3a-c, we use $\qp_B$ and $\qm_A$ for $F$. In Fig. 3d-i, we use $F=\pi\qm_A\qm_B$ to generate the background colors. The same choices are used in Fig. 6, though in different coordinates.

To display the overlaid vector field, however, requires a more involved formula. We use the exact same reactive current formula as in the supplement of \cite{Strahan2020}, but repeat it here for reference. The projected current is defined as
\begin{align}
    \bj_{AB}^{\by}(\vy)=\int\bj_{AB}(\vx)\cdot\nabla\by(\vx)\delta\big(\by(\vx)-\vy\big)\,d\vx
\end{align}
In the discretized $\vy$ space, this leads to the discretized projected current:
\begin{align}
    \bj_{AB}^\by(\vy)\approx\frac1{2\Delta t}\sum_{n=1}^N\chom(\bx_n(0))&\bigg[\ind_{d\vy}(\bx_n(0))\qm_A(\bx_n(0))\qp_B(\bx_n(\theta_n))\frac{\by(\bx_n(\theta_n))-\by(\bx_n(0))}{\theta_n}\\
    +\ind_{d\vy}&(\bx_n(\Delta t))\qm_A(\bx_n(\wt{\theta}_n))\qp_B(\bx_n(\Delta t))\frac{\by(\bx_n(\Delta t))-\by(\bx_n(\wt{\theta}_n))}{\Delta t-\wt{\theta}_n}\bigg]
\end{align}
where $\theta_n$ and $\wt{\theta}_n$ are the ``first-entry times'' to $D=(A\cup B)\cm$ in the $n$th trajectory with time running forward and backward, respectively. 

We use a similar formula to display the composites in Figs. 4, 5 and the stochastic tendencies in Fig. 7. Suppose we have a progress coordinate $f$ evaluated on all the data points. The composite SSW at a given level $f_0\pm\,df$ is approximated
\begin{align}
	\bigg\{\bx_n(t): |f(\bx_n(t))-f_0|<df\text{, weighted by }\qm_A(\bx_n(t))\qp_B(\bx_n(t))\der\pi\mu(\bx_n(0))\bigg\}
\end{align}
where $t$ can range from 0 to 20 days, the length of the short trajectory. In Fig. 4b, we use $-\tbd_B$ as the progress coordinate $f$, spanning the range from $-80$ days to zero with a step size of 5 days and a tolerance of $df=7.5$ days, except for the final level where the tolerance is reduced to 1.5 days. Smaller tolerances should be possible with a larger dataset, but in our case produced artifacts in the plot due to limited data. In Fig. 4c, we use $\qp_B$ as the progress coordinate $f$, spanning the range from 0 to 1 with a step size of 0.05 and a tolerance of 0.075, except for the final level where the tolerance is reduced to 0.015.

The stochastic tendencies shown in the bottom row of Fig. 7 are found in a similar way. For a given observable, such as $\gramps$, its stochastic tendency $\lcal_{AB}\gramps$ is estimated at a single point in each trajectory, in particular day $t_1=5$ days out of 20, as follows. First, we set $t_0=0$ days, or else the last-exit time from $A\cup B$ between days 0 and 5. Second, we set $t_2=10$ days, or else the first-entry time to $A\cup B$ between days 5 and 10. Then
\begin{align}
	\lcal_{AB}\gramps(\bx_n(t_1))\approx\frac{1}{2\qm_A(t_1)\qp_B(t_1)}\bigg[\frac{\gramps(\bx_n(t_2))-\gramps(\bx_n(t_1))}{t_2-t_1}\qm_A(t_1)\qm_B(t_2)-\frac{\gramps(\bx_n(t_1))-\gramps(\bx_n(t_0))}{t_1-t_0)}\qm_A(t_0)\qm_B(t_1)\bigg]
\end{align}
and we estimate the mean stochastic tendency across a surface of constant $\qp_B$ by 
the usual method of estimating ergodic averages, i.e., Eq.~\eqref{eq:discrete_ergodic_average}

\subsection{Algorithmic parameters}
Having sketched the general numerical procedure, we now provide the exact parameters used here, which are similar to those in \cite{Finkel2021learning}. We use $N=3\times10^5$ trajectories, each of length $\Delta t=20$ days, with a sampling interval of 1 day. The initial conditions are resampled from a long ($\times10^6$-day) control simulation to be uniformly distributed on the space $(|\Psi|(30\,\mathrm{km}),U(30\,\mathrm{km}))$. With a more complex, expensive model, we cannot rely on a control simulation to seed the initial points, but here we focus on TPT and DGA as a proof of concept rather than optimizing the numerical procedure. We use  $M=1500$ basis functions defined as indicators on a partition induced by $K$-means clustering on $\{\bx_n(0)\}$. The clustering is hierarchical so that the cluster size does not become too imbalanced. 

There are many potential directions for methodological improvement. In an expensive model without the ability to run a long control simulation, we should use a splitting and killing method to seed the initial trajectories across state space. Moreover, we could perform DGA repeatedly with new data seeded at each iteration in areas of high sensitivity. The choice of basis function can also powerfully affect DGA's performance. Indicator functions are advantageous in producing a \emph{bona fide} Markov matrix and guaranteeing a maximum principle for the committor probabilities. However, smooth and/or global basis functions have in some cases found to be more efficient at capturing the structure of the committor with fewer basis elements \cite{dga,Strahan2020}.

\section{Minimum-action method}
\label{sec:si_minimum_action_method} 

To compute the minimum-action paths, we use a completely discrete approach for simplicity and to accommodate the low-rank nature of the stochastic forcing. Heuristically, we wish to find the most probable path connecting $A$ and $B$, which we take as the mode of the (discretized) path density over the distribution of paths from $A$ to $B$. For concreteness, fix $\vx(0)=\vx_0\in A$ and a time horizon $T$ discretized into $K$ intervals, with a timestep $\delta t=T/K=0.005$ days. The discretized dynamics evolve according to the Euler-Maruyama method as
\begin{align}
    \vx(k\delta t)=\vx((k-1)\delta t)+\bm v\big(\vx((k-1)\Delta t)\big)\Delta t+\bm\sigma\bm\eta_k\sqrt{\delta t}
\end{align}
where $\bm\eta_k$ is a vector of i.i.d. unit normal samples, $\bm v$ is the deterministic drift, and $\bm\sigma\in\R^{d\times m}$ is the diffusion matrix, with a noise rank $m=3$ and a spatially smooth structure as defined in \cite{Finkel2021learning}. In the classical minimum-action approach, $\bm\sigma$ is assumed to be a $d\times d$ invertible matrix, and the probability density of a path $(\vx_0,\hdots,\vx_K)$ (where $\vx_k=\vx(k\delta t)$) is \begin{align}       
    \prod_{k=1}^K\ncal&\Big(\vx_k\Big|\vx_{k-1}+\bm v(\vx_{k-1})\delta t,\sigma\sigma\tr\delta t\Big)\\    &=\prod_{k=1}^K\frac1{(2\pi\delta t)^{dK/2}(\det\sigma)^K}\times\\
    &\exp\bigg\{-\frac12\Big(\vx_k-\vx_{k-1}-\bm v(\vx_{k-1})\delta t\Big)\tr\frac{(\sigma\sigma\tr)\inv}{\delta t}\Big(\vx_k-\vx_{k-1}-\bm v(\vx_{k-1})\delta t\Big)\bigg\}\\ 
    &\propto\exp\bigg\{-\frac{\delta t}2\sum_{k=1}^K\bigg(\frac{\vx_k-\vx_{k-1}}{\delta t}-\bm v(\vx_{k-1})\bigg)\tr(\sigma\sigma\tr)\inv\bigg(\frac{\vx_k-\vx_{k-1}}{\delta t}-\bm v(\vx_{k-1})\bigg)\bigg\}\\    &\sim\exp\bigg\{-\frac12\int_0^T\Big[\dot{\vx}(t)-\bm v(\vx(t))\Big](\sigma\sigma\tr)\inv\Big[\dot{\vx}(t)-\bm v(\vx(t))\Big]\,dt\bigg\}\text{ as }\delta t\to0
\end{align}
and the problem becomes to minimize the quadratic form in the argument of the exponential, which is the Freidlin-Wentzell action functional, subject to the constraint $\vx_K\in B$. However, because we stir the wind field with smooth spatial forcing in only $m\ll d$ wavenumbers, $\sigma$ is low-rank and thus $\bm\sigma\bm\sigma\tr$ is singular. Given any realized path $(\vx_0,\hdots,\vx_K)$, there may be no possible underlying forcing $\bm\eta_k$ that could have produced it under our noise model. So the obvious optimization strategy of fixing $\vx_0$ and $\vx_K$ and varying the steps in between may lead to impossible paths. For this reason, we perform optimization in the space of perturbations, and ensure that every step of the optimization is realizable under our noise model. This is a strategy we adopt from the cyclogenesis model \cite{Plotkin2019}. The result will be a simpler, convex objective function at the expense of a more complicated constraint. The probability density of a particular forcing sequence $(\bm\eta_1,\hdots,\bm\eta_K)$ is given by
\begin{align}  
    \prod_{k=1}^K\frac1{(2\pi)^{m/2}}\exp\bigg(-\frac12\bm\eta_k\tr\bm\eta_k\bigg)
    =\frac1{(2\pi)^{mK/2}}\exp\bigg(-\frac12\sum_{k=1}^K\bm\eta_k\tr\bm\eta_k\bigg)
\end{align}
The objective inside the exponential is now a simple quadratic in perturbation space which can be easily differentiated with respect to those perturbations. The constraint, meanwhile, takes the form of a complicated iterated function. Define the flow map $F(\vx)=\vx+\bm v(\vx)\delta t$ as the deterministic part of the timestep, so $\vx_k=F(\vx_{k-1})+\bm\sigma\bm\eta_k\sqrt{\delta t}$. In terms of $F$, the endpoint has to be written as a recursive function
\begin{align}
    \vx_K&=F(\vx_{K-1})+\sigma\bm\eta_K\sqrt{\delta t}\\    \vx_{K-1}&=F(\vx_{K-2})+\sigma\bm\eta_{K-1}\sqrt{\delta t}\\    &\vdots\\    \vx_1&=F(\vx_0)+\sigma\bm\eta_1\sqrt{\delta t}    
\end{align}
We impose the end constraint by adding to the action a penalty $\Phi(\vx_K)=\text{dist}(\vx_K,B)$, a function which linearly increases with distance to $B$. The full optimization problem is
\begin{align}   
    \min_{\bm\eta}&\bigg\{\frac1{2K}\sum_{k=1}^K\bm\eta_k\tr\bm\eta_k+\alpha\Phi(\vx_K)\bigg\}\\
    &\vx_0\in A\text{ is fixed}\\    &\vx_k=F(\vx_{k-1})+\bm\sigma\bm\eta_k\sqrt{\delta t}\text{ for }k=1,\hdots,K
\end{align}
Here $\alpha$ is a weight which can be increased to harden the end constraint. We divide by $K$ so that the path action does not overwhelm the endpoint penalty as $K\to\infty$. (This makes the sum converge to an integral.) We set $\vx_0$ to be the fixed point $\mathbf{a}\in A$ when finding the least-action path from $A$ to $B$ and the fixed point $\mathbf{b}\in B$ when finding the least-action path from $B$ to $A$. We used the L-BFGS method as implemented in \texttt{scipy}, with a maximum of 10 iterations. We differentiate $\Phi(x_K)$ with respect to $\bm\eta_k$ using knowledge of the adjoint model, with a backward pass through the path to compute each gradient. At each descent step, we refine the stepsize with backtracking line search. One way to guarantee the end constraint is ultimately satisfied is to gradually increase $\alpha$ and lengthen $T$; however, we found it sufficient to fix $\alpha=1.0$ and $T=100$, in keeping with the typical observed transit time. We have kept the algorithm simple, not devoting too much effort to finding the global optimimum over all time horizons, as we only care for a qualitative assessment to compare with results of TPT.

\section{Enstrophy budget derivation}
\label{sec:si_enstrophy_budget}

Here we derive an enstrophy budget from the Holton-Mass model. For completeness, we also provide a derivation of the original Holton-Mass model, starting from Eqs. (1) and (6) of \cite{holton_mass} and deriving their Eqs. (9) and (10), equivalent to Eqs. (3) of our main text. This provides some helpful context for the enstrophy budget.

\subsection{Holton-Mass projected model}
This section inserts the ansatz (Eqs. 7 of \cite{holton_mass})
\begin{subequations}
	\label{eq:si_ansatz}
	\begin{align}
		\psi'(x,y,z,t)&=\re\{\Psi(z,t)e^{z/2H}e^{ikx}\sin\ell y\}\\
		u(x,y,z,t)&=U(z,t)\sin\ell y
	\end{align}
\end{subequations}
into the wave-mean flow model. The equations are nonlinear, involving products of pairs of terms containing $\sin\ell y$, which projects onto wavenumber $2\ell$. Following \cite{holton_mass}, we approximate $\sin^2\ell y$ by $\eps\sin\ell y$, where $\eps=8/(3\pi)$, to close the equations in terms of a single wavenumber. We will also make extensive use of the identity
\begin{align}
	e^{cz}\partial_z\big(e^{-cz}f(z)\big)=(\partial_z-c)f(z)
\end{align}

\subsubsection{Wave equation}
We start with the quasi-geostrophic potential vorticity equation, Eq. (1) of \cite{holton_mass}, which governs the tendency of the streamfunction $\Psi$:
\begin{subequations}
	\label{eq:si_wave_eqn}
	\begin{align}
		0&=(\partial_t+\ov u\partial_x)q'+\partial_y\ov q\partial_x\psi'+\fn e^{z/H}\partial_z(e^{-z/H}\alpha\partial_z\psi')
	\end{align}
	where 
	\begin{align}
		q'&=\nabla^2\psi'+\fn e^{z/H}\partial_z(e^{-z/H}\partial_z\psi')\\
		\partial_y\ov q&=\beta-\partial_y^2\ov u-\fn e^{z/H}\partial_z(e^{-z/H}\partial_z\ov u)
	\end{align}
\end{subequations}
We deal with the three terms one at a time. All three terms are linear in $\psi'$, and so we can take the real part outside. The first term---the material derivative of PV following the mean flow---is 
\begin{subequations}
	\begin{align}
		(\partial_t+\ov u\partial_x)q'&=\re\bigg\{(\partial_t+Uik\sin\ell y)\bigg[-(k^2+\ell^2)\Psi e^{z/2H}
		+\fn e^{z/H}\partial_z\Big(e^{-z/H}\partial_z(\Psi e^{z/2H})\Big)\bigg]e^{ikx}\sin\ell y\bigg\}\\
		&\approx\re\bigg\{(\partial_t+Uik\eps)\bigg[-(k^2+\ell^2)\Psi e^{z/2H}
		+\fn e^{z/2H}e^{z/2H}\partial_z\bigg(e^{-z/2H}e^{-z/2H}\partial_z(e^{z/2H}\Psi)\bigg)\bigg]e^{ikx}\sin\ell y\bigg\}\\
		&=\re\bigg\{(\partial_t+Uik\eps)\bigg[-(k^2+\ell^2)\Psi e^{z/2H}
		+\fn e^{z/2H}\bigg(\partial_z-\frac{1}{2H}\bigg)\bigg(\partial_z+\frac{1}{2H}\bigg)\Psi\bigg]e^{ikx}\sin\ell y\bigg\}\\		
		&=\re\bigg\{(\partial_t+Uik\eps)\bigg[-(k^2+\ell^2)\Psi
		+\fn\bigg(\partial_z^2-\frac{1}{4H^2}\bigg)\Psi\bigg]e^{z/2H}e^{ikx}\sin\ell y\bigg\}\\	
	\end{align}
\end{subequations}

The second term---meridional advection of PV due to eddies---is 
\begin{align}
	\partial_y\ov q\partial_x\psi'&=\re\bigg\{\bigg[\beta+\ell^2U\sin\ell y-\fn\bigg(U_{zz}-\frac1HU_z\bigg)\sin\ell y\bigg]ik\Psi e^{z/2H}e^{ikx}\sin\ell y\bigg\}\\
	&\approx\re\bigg\{ik\Psi\bigg[\beta+\eps\bigg(\ell^2U+\fn\Big(\frac1HU_z-U_{zz}\Big)\bigg)e^{z/2H}e^{ikx}\sin\ell y\bigg\}
\end{align}

The third term---radiative cooling---is
\begin{align}
	\fn e^{z/H}\partial_z(e^{-z/H}\alpha\partial_z\psi')
	&=
	\re\bigg\{\fn\bigg[e^{z/2H}e^{z/2H}\partial_z\bigg(e^{-z/2H}\alpha e^{-z/2H}\partial_z\big(e^{z/2H}\Psi\big)\bigg)\bigg]e^{ikx}\sin\ell y\bigg\}\\
	&=\re\bigg\{\fn\bigg(\partial_z-\frac1{2H}\bigg)\bigg[\alpha\bigg(\partial_z+\frac1{2H}\bigg)\Psi\bigg]e^{z/2H}e^{ikx}\sin\ell y\bigg\}
\end{align}
We now combine the three terms. Stipulating they add to zero for all $x$ actually means we can drop the ``real part'' on the outside. This is because for any complex function $f(x)=g(x)+ih(x)$ for real $g$ and $h$,
\begin{align}
	\re\{f(x)e^{ikx}\}=\re\{\big(g(x)+ih(x)\big)(\cos kx+i\sin kx)\}=g(x)\cos kx-h(x)\sin kx
\end{align}
By orthogonality of $\sin(kx)$ and $\cos(kx)$ on the domain, both $g$ and $h$ must vanish identically if the left-hand side is zero. Hence, we recover the projected QGPV equation, Eq. (9) of \cite{holton_mass}:
\begin{align}
	0=(\partial_t+ik\eps U)&\bigg[-\bigg(k^2+\ell^2+\fn\frac1{4H^2}\bigg)+\fn\partial_z^2\bigg]\Psi \\
	+ik\Psi&\bigg[\beta+\eps\ell^2U+\eps\fn\bigg(\frac1HU_z-U_{zz}\bigg)\bigg] \nonumber\\
	+\fn&\bigg(\partial_z-\frac1{2H}\bigg)\bigg[\alpha\bigg(\partial_z+\frac1{2H}\bigg) \Psi\bigg] \nonumber
\end{align}
We take a further step and non-dimensionalize for a more compact expression. We select a horizontal length scale $L=2.5\times10^5$ m, a vertical length scale $H=7\times10^3$ m (the same as the $H$ in the equation), and a time scale $T=1$ day $=86400$ s. Thus rescaling all variables,
\begin{align}
	0=\frac1T(\partial_t+ik\eps U)&\bigg[-\bigg(\frac{k^2+\ell^2}{L^2}+\fn\frac1{4H^2}\bigg)+\fn\frac1{H^2}\partial_z^2\bigg]\frac{L^2}{T}\Psi\\
	&+\frac{L}{T}ik\Psi\bigg[\frac{\beta}{LT}+\frac{\eps\ell^2U}{LT}+\eps\fn\bigg(\frac{L}{H^2T}U_z-\frac{L}{H^2T}U_{zz}\bigg)\bigg] \nonumber\\
	&+\fn\frac1H\bigg(\partial_z-\frac12\bigg)\bigg[\frac{\alpha}{HT}\bigg(\partial_z+\frac12\bigg)\frac{L^2}{T}\Psi\bigg] \nonumber
\end{align}
We now define the dimensionless group $\gcal^2:=H^2N^2/(L^2f_0^2)$ and multiply through by $T^2\gcal^2$ and obtain
\begin{align}
	\label{eq:si_nondim_qgpv}
	(\partial_t+ik\eps U)&\bigg[-\gcal^2(k^2+\ell^2)-\frac14+\partial_z^2\bigg]\Psi\\
	&+ik\Psi\bigg[\gcal^2\beta+\eps\big(\gcal^2\ell^2U+U_z-U_{zz}\big)\bigg]\nonumber \\
	&\ \ =-\bigg(\partial_z-\frac12\bigg)\bigg[\alpha\bigg(\partial_z+\frac12\bigg)\Psi\bigg] \nonumber
\end{align}
exactly as in Eq. (3) of the main paper. 

\subsubsection{Mean-flow equation}

The PDE for $\ov u$ is specified in Eq. (6) of \cite{holton_mass}:
\begin{align}
	\partial_t&\bigg[\partial_y^2\ov u+\fn e^{z/H}\partial_z\big(e^{-z/H}\partial_z\ov u\big)\bigg]\\
	&=-\fn e^{z/H}\partial_z\big[\alpha e^{-z/H}\partial_z(\ov u-u_R)\big]\\
	&+\fn\partial_y^2\big[e^{z/H}\partial_z\big(e^{-z/H}\ov{\partial_x\psi'\partial_z\psi'})\big]
\end{align}
Putting in the ansatz for $U$ renders the first two terms trivial. The left-hand side becomes
\begin{align}
	\partial_t\bigg[-\ell^2U+\fn\bigg(U_{zz}-\frac1HU_z\bigg)\bigg]\sin\ell y
\end{align}
and setting $u_R(z)=U^R(z)\sin\ell y$, the first term on the right becomes 
\begin{align}
	-\fn e^{z/H}\partial_z\big[\alpha e^{-z/H}\partial_z(U-U^R)\big]
\end{align}
The third term is more involved due to the zonal correlation term. However, we can simplify using the rule for zonal correlations from Eq. (25) of the main text, which is derived more fully here. Let $\psi_1'=\re\{\Psi_1e^{ikx}\}$ and $\psi_2'=\re\{\Psi_2e^{ikx}\}$, where $\Psi_1,\Psi_2$ are two complex numbers independent of $x$ with real parts $X_1,X_2$ and imaginary parts $Y_1,Y_2$ respectively. Their zonal eddy correlation is
\begin{align}
	\ov{\psi_1'\psi_2'}&=\ov{\re\{\Psi_1e^{ikx}\}\re\{\Psi_2e^{ikx}\}}\\
	&=\ov{(X_1\cos kx-Y_1\sin kx)(X_2\cos kx-Y_2\sin kx)}\\
	&=X_1X_2\ov{\cos^2kx}-(X_1Y_2+Y_1X_2)\ov{\cos kx\sin kx}+Y_1Y_2\ov{\sin^2kx}\\
	&=\frac12(X_1X_2+Y_1Y_2)\\
	&=\frac12\re\{(X_1-iY_1)(X_2+iY_2)\}=\frac12\re\{\Psi_1^*\Psi_2\}
\end{align}
Using this simple rule, we have
\begin{align}
	\ov{\partial_x\psi'\partial_z\psi'}&=\frac12\re\Big\{\Big(ik\Psi e^{z/2H}\sin\ell y\Big)^*\partial_z\Big(\Psi e^{z/2H}\sin\ell y\Big)\Big\}\\
	&\approx\frac12k\eps\,\re\bigg\{-i\Psi^*e^{z/2H}\bigg(\Psi_z+\frac1{2H}\Psi\bigg)e^{z/2H}\bigg\}\sin\ell y\\
	&=\frac12k\eps\,\im\{\Psi^*\Psi_z\}e^{z/H}\sin\ell y
\end{align}
where we have used that $\re\{-i\Psi^*\Psi\}=0$. The second term on the right-hand side of the mean-flow equation is then
\begin{align}
	-\fn\ell^2 e^{z/H}\partial_z\bigg(\frac12k\eps\,\im\{\Psi^*\Psi_z\}\bigg)\sin\ell y
	&=-\fn\frac{\eps k\ell^2}{2}e^{z/H}\im\{\Psi^*\Psi_{zz}\}\sin\ell y
\end{align}
where we have used $\im\{\Psi_z^*\Psi_z\}=0$. Putting these terms together, dropping the $\sin\ell y$, and negating each term, we reproduce Eq. (10) of \cite{holton_mass}:
\begin{align}
	\partial_t\bigg[\ell^2U+\fn\bigg(\frac1HU_z-U_{zz}\bigg)\bigg]
	=\fn e^{z/H}\partial_z\big[\alpha e^{-z/H}\partial_z(U-U^R)\big]+\fn\frac{\eps k\ell^2}{2}e^{z/H}\im\{\Psi^*\Psi_{zz}\}
\end{align}
Let us non-dimensionalize this equation as we did for the mean-flow equation with the scales $L$, $H$, and $T$. 
\begin{align}
	\frac1T\partial_t\bigg[\frac{1}{LT}\ell^2U+\fn\frac{L}{H^2T}\big(U_z-U_{zz}\big)\bigg]
	&=\fn e^z\frac1H\partial_z\bigg[\frac{\alpha}{T}e^{-z}\frac{L}{HT}\partial_z(U-U^R)\bigg]+\fn\frac1{L^3}\frac{\eps k\ell^2}{2}e^z\frac{L^4}{T^2H^2}\im\{\Psi^*\Psi_{zz}\}
\end{align}
Multiplying through by $LT^2\gcal^2$,
\begin{align}
	\label{eq:si_nondim_meanflow}
	\partial_t\big[\gcal^2\ell^2U+U_z-U_{zz}\big]&=e^z\partial_z\big[\alpha e^{-z}\partial_z(U-U^R)\big]+\frac{\eps k\ell^2}{2}e^z\im\{\Psi^*\Psi_{zz}\}
\end{align}
Notice that the first term in brackets, $\gcal^2\ell^2U+U_z-U_{zz}$, also appears in the nonlinear coupling on the right-hand side of the QGPV equation~\eqref{eq:si_nondim_qgpv}. To obtain more symmetry between the wave and mean-flow equations, we add $(\gcal^2/\eps)\partial_t\beta$ (which is zero) to the left-hand side of Eq.~\eqref{eq:si_nondim_meanflow}, and then multiply through by $2/(\eps\ell^2)$:
\begin{align}
	\partial_t\bigg[\frac{\gcal^2}{\eps}\beta+\gcal^2\ell^2U+U_z-U_{zz}\bigg]&=e^z\partial_z\big[e^{-z}\alpha\partial_z(U-U^R)\big]+\frac{\eps k\ell^2}{2}e^z\im\{\Psi^*\Psi_{zz}\}\\
	\frac{2}{(\eps\ell)^2}\partial_t\big[\gcal^2\beta+\eps\big(\gcal^2\ell^2U+U_z-U_{zz}\big)\big]&=\frac2{\eps\ell^2}e^z\partial_z\big[e^{-z}\alpha\partial_z(U-U^R)\big]+ke^z\,\im\{\Psi^*\Psi_{zz}\}
\end{align}
This is Eq. (3)b of our main text. 

In the next section, we derive an enstrophy budget directly from these projected equations.

\subsection{Enstrophy budget}

The enstrophy budget starts with the projected wave-mean flow interaction equations, repeated below with the following abbreviations:
\begin{itemize}
	\item $\beta_e=\gcal^2\beta+\eps(\gcal^2\ell^2U+U_z-U_{zz})$. \cite{holton_mass} used the same symbol to denote the corresponding quantity in the PDE, $\beta-\partial_y^2\ov u-e^{z/H}\partial_z[e^{-z/H}\partial_z\ov u]$, which physically represents meridional gradient of zonal-mean PV. Here, we use $\partial_y\ov q$ for the PDE version, and $\beta_e$ strictly for the projected version of the equation.  
	\item $\delta=\gcal^2(k^2+\ell^2)+\frac14$. This constant term appears in the expression for QGPV, arising from the three-dimensional Laplacian. 
	\item $F_q=ke^z\im\{\Psi^*\Psi_{zz}\}$. This term represents the eddy meridional PV flux. 
	\item $R=\frac2{\eps\ell^2}e^z\partial_z\big[e^{-z}\alpha\partial_z(U-U^R)\big]$. This term represents thermal relaxation of the wind field toward the radiative wind $U^R$.  
\end{itemize}

With these abbreviations, the wave and mean-flow equations are 
\begin{subequations}
	\begin{align}
		\frac{2}{(\eps\ell)^2}\partial_t\beta_e-F_q&=R \label{eq:meanflow_eqn_abbrv}\\
		(\partial_t+ik\eps U)\big(-\delta+\partial_z^2\big)\Psi
		+ik\Psi\beta_e
		&=-\bigg(\partial_z-\frac12\bigg)\bigg[\alpha\bigg(\partial_z+\frac12\bigg)\Psi\bigg]  \label{eq:wave_eqn_abbrv}
	\end{align}	
\end{subequations}

Three steps will give us the enstrophy budget. 

\begin{description} 

\item[Step (i)] Multiply Eq.~\eqref{eq:meanflow_eqn_abbrv} by $\beta_e$.

\begin{subequations}
	\begin{align}
		\frac{1}{(\eps\ell)^2}\partial_t\big(\beta_e^2\big)-F_q\beta_e&=R\beta_e\\
		\partial_t\gramps-F_q\beta_e&=R\beta_e \label{eq:si_gramps_evolution}\\
		\text{where }\gramps&:=\bigg(\frac{\beta_e}{\eps\ell}\bigg)^2
	\end{align}
\end{subequations}

\item[Step (ii)] Multiply Eq.~\eqref{eq:wave_eqn_abbrv} by $(-\delta+\partial_z^2)\Psi^*$, and then take the real part. This corresponds to multiplying by $q'$ and taking the zonal mean, as in the standard EP relation. 

\begin{subequations}
	\begin{align}
		\re\Big\{\big[(-\delta+\partial_z^2)\Psi^*\big](\partial_t+ik\eps U)(-\delta+\partial_z^2)\Psi\Big\}
		&+k\beta_e\re\Big\{i\big[(-\delta+\partial_z^2)\Psi^*\big]\Psi\Big\} \\
		&=-\re\bigg\{\big[(-\delta+\partial_z^2)\Psi^*\big]\bigg(\partial_z-\frac12\bigg)\bigg[\alpha\bigg(\partial_z+\frac12\bigg)\Psi\bigg]\bigg\}
	\end{align}
\end{subequations}
Let us simplify the terms one at a time. The first term on the left can be easily simplified by applying the identity
\begin{align}
	\partial_t\Big(\frac12|f(t)|^2\Big)=\frac12\partial_t(f^*f)=\frac12(f_t^*f+f^*f_t)=\re\{f_t^*f\}
\end{align}
to $f=(-\delta+\partial_z^2)\Psi^*$:
\begin{align}
	\re\Big\{&\big[(-\delta+\partial_z^2)\Psi^*\big](\partial_t+ik\eps U)(-\delta+\partial_z^2)\Psi\Big\}\\
	&=\re\bigg\{\partial_t\bigg(\frac12\big|(-\delta+\partial_z^2)\Psi\big|^2\bigg)\bigg\}
	+k\eps U\,\re\Big\{i\big|(-\delta+\partial_z^2)\Psi\big|^2\Big\}\\
	&=\partial_t\bigg(\frac12\big|(-\delta+\partial_z^2)\Psi\big|^2\bigg)
\end{align}
The second term becomes
\begin{align}
	k\beta_e\re\big\{-\delta i\Psi^*\Psi+i\Psi^*_{zz}\Psi\big\}=-k\beta_e\im\{\Psi_{zz}^*\Psi\}=k\beta_e\im\{\Psi^*\Psi_{zz}\}=F_q\beta_ee^{-z}
\end{align}
where in the last step we recognized the same term from the mean-flow equation. The right-hand side cannot be very simplified, but we expand term by term for an alternative expression:
\begin{align}
	-\re&\bigg\{\big[(-\delta+\partial_z^2)\Psi^*\big]\bigg(\partial_z-\frac12\bigg)\bigg[\alpha\bigg(\partial_z+\frac12\bigg)\Psi\bigg]\bigg\}\\
	&=-\re\bigg\{\big(-\delta\Psi^*+\Psi_{zz}^*\big)\bigg(\partial_z-\frac12\bigg)\bigg(\alpha\Psi_z+\frac\alpha2\Psi\bigg)\bigg\}\\
	&=-\re\bigg\{\big(-\delta\Psi^*+\Psi_{zz}^*\big)\bigg[\alpha\Psi_{zz}+\alpha_z\Psi_z+\bigg(\frac{\alpha_z}{2}-\frac\alpha4\bigg)\Psi\bigg]\bigg\}\\
	&=-\alpha|\Psi_{zz}|^2-\alpha_z\re\{\Psi_{zz}^*\Psi_z\}-\bigg(\frac{\alpha_z}{2}-\frac\alpha4\bigg)\re\{\Psi_{zz}^*\Psi\}\\
	&\hspace{1.0cm}+\delta\bigg[\alpha\,\re\{\Psi^*\Psi_{zz}\}+\alpha_z\re\{\Psi^*\Psi_z\}+\bigg(\frac{\alpha_z}{2}-\frac\alpha4\bigg)|\Psi|^2\bigg]\\
	&=\delta\bigg(\frac{\alpha_z}{2}-\frac\alpha4\bigg)|\Psi|^2+\delta\alpha_z\re\{\Psi^*\Psi_z\}\\
	&\hspace{1.0cm}+\bigg(\delta\alpha+\frac\alpha4-\frac{\alpha_z}{2}\bigg)\re\{\Psi_{zz}^*\Psi\}-\alpha_z\re\{\Psi_{zz}^*\Psi_z\}-\alpha|\Psi_{zz}|^2
\end{align}

Finally, we recombine all three terms and multiply through by $e^z$ for symmetry with the mean-flow equation.
\begin{align}
	\partial_t\ecal
	+F_q\beta_e
	&=D  \label{eq:si_enstrophy_evolution}\\
	\text{where}&\\
	\ecal&=\frac12e^z\Big|\big(-\delta+\partial_z^2\big)\Psi\Big|^2 \\
	D&=-\re\bigg\{e^z\big[\big(-\delta+\partial_z^2)\Psi^*\big)\big]\bigg(\partial_z-\frac12\bigg)\bigg[\alpha\bigg(\partial_z+\frac12\bigg)\Psi\bigg]\bigg\}
\end{align}

\item[Step (iii)]: Add together the evolution equations~\eqref{eq:si_gramps_evolution} and~\eqref{eq:si_enstrophy_evolution} for $\gramps$ and $\ecal$. 

\begin{align}
	\partial_t\ecal&=D-F_q\beta_e\\
	\partial_t\gramps&=R\beta_e+F_q\beta_e\\
	\therefore\partial_t(\gramps+\ecal)&=R\beta_e+D
\end{align}
\end{description}

The advantage of this arrangement is the left-hand side terms are quadratic in $\Psi$ and $U$ respectively, while the right-hand side terms contain all the information about radiative damping via the Newtonian cooling coefficient $\alpha(z)$. In the absence of such dissipation, therefore, $\gramps+\ecal$ would be a conserved quantity. 

\bibliographystyle{apalike}
\bibliography{tpt_references}